\pdfoutput=1

\documentclass[11pt]{article}

\usepackage[margin=1in]{geometry}
\usepackage[english]{babel}
\usepackage[compact]{titlesec}
\usepackage[T1]{fontenc}

\usepackage{authblk} % necessary for some reason (5/15/25)

\usepackage[svgnames,xcdraw,table]{xcolor}
\definecolor{myBlue}{rgb}{0.1,0.1,0.8}
\definecolor{DarkGreen}{rgb}{0.1,0.5,0.1}
\usepackage{hyperref}
\hypersetup{
	colorlinks=true,
	linkcolor=red,%color of internal links
	urlcolor=DarkGreen,%color of external links
	citecolor=blue%color of links to bibliography
}
\usepackage[hyperpageref]{backref}
\renewcommand*{\backref}[1]{}
\renewcommand*{\backrefalt}[4]{%
    \ifcase #1 (Not cited.)%
    \or        (Cited on page~#2)%
    \else      (Cited on pages~#2)%
    \fi}

\usepackage{times}
\usepackage{soul}
\usepackage{url}
\usepackage[utf8]{inputenc}
\usepackage{caption}
\usepackage{graphicx}
\usepackage{amsmath}
\usepackage{amsthm}
\usepackage{booktabs}
\usepackage{algorithm}
\usepackage{algorithmic}
\usepackage[switch]{lineno}

\usepackage{natbib}

\newcommand{\EF}[1]{\ifstrempty{#1}{\textrm{\textup{EF}}}{\textrm{\textup{EF{$#1$}}}}}
\newcommand{\HEF}[1]{\ifstrempty{#1}{\textrm{\textup{HEF}}}{\textrm{\textup{HEF-{$#1$}}}}}

\usepackage{amsfonts}
\usepackage{amsthm}
\usepackage[inline]{enumitem}%[inline]
\usepackage{etoolbox}
\usepackage{makecell}
\usepackage{mathtools}
\usepackage{multirow}
\usepackage{thmtools}
\usepackage{thm-restate}
\usepackage{wasysym}
\usepackage[capitalise,noabbrev]{cleveref}
\Crefname{remark}{Remark}{Remarks}
\Crefname{rmk}{Remark}{Remarks}
\Crefname{dfn}{Definition}{Definitions}
\Crefname{thm}{Theorem}{Theorems}
\Crefname{cor}{Corollary}{Corollaries}
\Crefname{lem}{Lemma}{Lemmas}
\Crefname{examplex}{Example}{Examples}
\Crefname{prop}{Proposition}{Propositions}

\usepackage{tikz}
\usetikzlibrary{shapes,fit,calc}

\colorlet{mygray}{gray!40}

\newcommand*\circled[1]{\tikz[baseline=(char.base)]{
		\node[shape=circle,draw,inner sep=1pt] (char) {#1};}}%fill=green!30

\newcommand*\circledcompact[1]{\tikz[baseline=(char.base)]{
		\node[shape=circle,draw,inner sep=-1pt] (char) {#1};}}%fill=green!30

\newcommand*\sqboxed[1]{\tikz[baseline=(char.base)]{
		\node[shape=rectangle,draw,inner sep=1pt] (char) {#1};}}%fill=green!30

\newcommand*\diamondboxed[1]{\tikz[baseline=(char.base)]{
		\node[shape=diamond,draw,inner sep=1pt] (char) {#1};}}%fill=green!30

\newcommand*\diamondboxedcompact[1]{\tikz[baseline=(char.base)]{
		\node[shape=diamond,draw,inner sep=0pt] (char) {#1};}}%fill=green!30

\declaretheoremstyle[
    bodyfont=\normalfont
]{normalstyle}

\declaretheorem[style=normalstyle,name=Definition]{dfn}

\declaretheorem[style=normalstyle,name=Example]{examplex}

\newcommand{\bn}{\mathbb{N}}
\DeclareRobustCommand{\Chi}{{\mathpalette\irchi\relax}}
\newcommand{\irchi}[2]{\raisebox{\depth}{$#1\chi$}} 
\newcommand{\sEF}[1]{\textrm{\textup{sEF{$#1$}}}}
\newcommand{\lbal}{\texttt{large}\texttt{-}\texttt{balanced}}
\newcommand{\lunbal}{\texttt{large}\texttt{-}\texttt{unbalanced}}
\newcommand{\mi}{\mathcal{I}}
\newcommand{\ns}{\text{ns}}

\newcommand{\sbal}{\texttt{small}\texttt{-}\texttt{balanced}}
\newcommand{\sunbal}{\texttt{small}\texttt{-}\texttt{unbalanced}}
\newcommand{\smallq}{\texttt{small}}
\newcommand{\largeq}{\texttt{large}}
\newcommand{\bal}{\texttt{balanced}}
\newcommand{\unbal}{\texttt{unbalanced}}
\newcommand{\stayopt}{\texttt{stay-is-opt}}
\newcommand{\swapopt}{\texttt{swap-is-opt}}

\usepackage{MnSymbol}%,bbding,pifont} % for \FiveStar

\usepackage[capitalise,noabbrev]{cleveref}

\title{Epistemic vs. Counterfactual Fairness in Allocation of Resources}

\author[1]{Hadi Hosseini}
\author[2]{Joshua Kavner}
\author[3]{Sujoy Sikdar}
\author[4]{Rohit Vaish}
\author[5]{Lirong Xia}
\affil[1]{Pennsylvania State University}
\affil[2]{Rensselaer Polytechnic Institute}
\affil[3]{Binghamton University}
\affil[4]{Indian Institute of Technology Delhi}
\affil[5]{Rutgers University}

\begin{document}

\maketitle

\begin{abstract}
Resource allocation is fundamental to a variety of societal decision-making settings, ranging from the distribution of charitable donations to assigning limited public housing among interested families. A central challenge in this context is ensuring fair outcomes, which often requires balancing conflicting preferences of various stakeholders. While extensive research has been conducted on theoretical and algorithmic solutions within the fair division framework, much of this work neglects the subjective perception of fairness by individuals. This study focuses on the fairness notion of envy-freeness (\EF{}), which ensures that no agent prefers the allocation of another agent according to their own preferences. While the existence of exact \EF{} allocations may not always be feasible, various approximate relaxations, such as counterfactual and epistemic \EF{}, have been proposed. Through a series of experiments with human participants, we compare perceptions of fairness between three widely studied counterfactual and epistemic relaxations of \EF{}. Our findings indicate that allocations based on epistemic \EF{} are perceived as fairer than those based on counterfactual relaxations. Additionally, we examine a variety of factors, including scale, balance of outcomes, and cognitive effort involved in evaluating fairness and their role in the complexity of reasoning across treatments. 
\end{abstract}

\section{Introduction}

Resource allocation is a pivotal concern in societal decision-making, attracting significant interest from disciplines as diverse as philosophy, economics, mathematics, and computer science.
It captures a wide range of application domains including distributing charitable donations to home shelters \citep{aleksandrov2015online}, assigning limited public housing to families and refugees \citep{andersson2018dynamic,ahani2024dynamic}, splitting rent among renters \citep{goldman2015spliddit} and job scheduling across distributed computing clusters \citep{isard2009quincy}.
A critical challenge in the allocation of resources among multiple stakeholders is ensuring \textit{fair} outcomes, necessitating the consideration of (often conflicting) preferences of participating entities (aka \emph{agents}).
While these problems have been extensively studied under the framework of \textit{fair division} in recent years, much of the focus has been on theoretical and algorithmic approaches (see e.g., \citet{M19fair,ALM+22algorithmic}, and \citet{ABF+22fair} for surveys concerning recent progress in this field). 
However, such approaches frequently neglect individuals' \textit{subjective} perception of fairness, which may diverge from their theoretical guarantees.\footnote{The perception of fairness has been recently studied in the context of loan decisions and within machine learning \citep{saxena2019fairness,srivastava2019mathematical}. These problems are fundamentally different from the current study as they are primarily concerned with bias in the data or prediction models.}

Our focus is on the fairness notion of \textit{envy-freeness} (EF), which requires that no agent prefers the bundle
% allocation or outcome
assigned to another agent when evaluated according to their own preferences \citep{foley1966resource}.
Among several plausible fairness notions---for example, those ensuring that each agent receives a fair share---envy-freeness is particularly compelling due to its reliance on pairwise \textit{intrapersonal} utility comparisons, eliminating the need for interpersonal comparisons. In other words, envy-freeness does not require identifying which agent derives the most benefit from a bundle of resources.
%each agent compares the outcome only with respect to his or her own preferences. 
% eliminates the need for identifying which agent derives the most benefit from a bundle of resources
% In addition, a substantial body of experimental studies in economics highlights its dominant role in human subject studies as a fairness criterion in resource allocation.
In addition, a substantial body of experimental studies in economics underscores its pivotal role as a fairness criterion in resource allocation
% , particularly in human subject studies 
\citep{herreiner2009envy,herreiner2010inequality}.

When dealing with scarce indivisible resources, \EF{} allocations do not always exist. When two families are both interested in a single 
% available public housing;
house, for instance, no \EF{} solution is possible. 
Furthermore, determining whether a resource allocation problem admits an \EF{} solution is known to be computationally intractable \citep{Lipton04:Approximately}.
These negative results have inspired a significant body of research aimed at developing approximate relaxations of envy-freeness.

% These negative results have motivated a 
% In light of the impossibility of guaranteeing \EF{} exactly, several relaxation of \EF{} have been proposed.

% \HH{
% Envy based notions are fundemnetal; lots of research in the past.
% Non-existence has given rise to approximations.
% Many approximations, unclear which is better with respect to preferences.
% }

The design of approximate fairness notions has given rise to two prominent schools of thought: epistemic and counterfactual envy-freeness.
The \textit{epistemic} approach focuses on the limited information that agents may possess about the overall allocation. In particular, \emph{envy-freeness up to $k$ hidden goods} (\HEF{k}) assumes agents have common information about how (and to whom) the goods are distributed except for a small subset of $k$ goods \citep{Hosseini2020:Fair}. 
Thus, agents have no envy given the information that is available to them. 
%
% \HEF{k} thus strikes a balance between the counterfactual removal of goods, as in \EF{1}, and \emph{epistemic envy-freeness}, which considers agents with different information about how goods are distributed \citep{ABC+18knowledge}.
%
In contrast, the \textit{counterfactual} approach centers on `hypothetical scenarios' which evaluate fairness based on 
% how outcomes would have differed had the 
allocations determined under
% been adjusted under 
different circumstances.
Specifically, a well-studied relaxation of envy-freeness is  \textit{envy-freeness up to one good} (\EF{1}). This notion is based on the \textit{counterfactual} thinking that any pairwise envy can be eliminated by the hypothetical removal of a \textit{single} good from the envied agent's bundle~\citep{Lipton04:Approximately}.\footnote{In the past decade, a myriad of counterfactual approximations have been proposed in the fair division literature. \EF{1} stands out because of its algorithmic simplicity and its clear implementation \citep{Lipton04:Approximately,B11combinatorial}.}
Despite their theoretical foundations, it remains unclear which approach is perceived as more desirable by humans. This leads to the following research question:

\begin{quote}\textit{
How do individuals perceive epistemic approximation of fairness compared to counterfactual approximations in allocation of resources?
}
\end{quote}

% We compare two prominent variants of \EF{}.
% First, \textit{envy-freeness up to one good} (\EF{1}) is based on the \textit{counterfactual} thinking that any pairwise envy can be eliminated by the hypothetical removal of a single good from the envied agent's bundle~\citep{B11combinatorial}.
% An \EF{1} allocation always exists 
% and can be computed in polynomial time \citep{Lipton04:Approximately}.
% Second, \emph{envy-freeness up to $k$ hidden goods} (\HEF{k})
% assumes agents have common information about an allocation except for a small subset of $k$ goods \citep{Hosseini2020:Fair}. Agents are envy-free given their available information. \HEF{k} thus strikes a balance between the counterfactual removal of goods, as in \EF{1}, and \emph{epistemic envy-freeness}, which considers agents with different information about how goods are distributed \citep{ABC+18knowledge}.

% \change{
\subsection{Our Contributions}

We study the perceived fairness of two variants of counterfactual envy-freeness (namely, \EF{1} and \emph{strong envy-freeness up to one good} (\sEF{1}) \citep{conitzer19:group}) compared to the epistemic notion of \HEF{} through a series of experiments with human subjects. 
%
% We focus on perceptions of envy, the specific aspect of fairness that underlies envy-based fairness notions, and estimate how likely it is for human subjects to experience envy given allocations that satisfy different relaxations of \EF{}. 
We conduct a study with 120 participants recruited through Amazon's Mechanical Turk platform. 
At a high level, our work is aligned with a large body of work in \textit{distributive justice} concerning fair outcomes (in contrast to \textit{procedural justice}, which concerns fair processes for determining outcomes).\footnote{We refer the reader to literature in social justice theory, e.g., \citet{adams1963towards} and \citet{rawls2004theory}. See \citet{tyler2002procedural}, \citet{rawls2004theory}, and \citet{Lee19:Justice} on procedural justice.}

\paragraph{\textbf{Framework and Fairness Measure.}}
We develop a novel empirical framework and a new approach for \textit{implicit} measurement of perceived fairness. 
% In our empirical study, 
Each participant is presented with a series of scenarios in which they take on the \textit{perspective} of an agent in a resource allocation instance with an initial allocation
that
satisfies one of three fairness properties: 
\sEF{1}, \EF{1}, or \HEF{k} (defined in Section  \ref{sec:model_defs}). 
Participants may either keep their given bundle or \emph{swap} it with the bundle of an agent of their choice. 
Our approach measures perceptions of fairness 
by evaluating whether an agent is envious of another's bundle.
If an agent is envious, then it is likely the agent would be
willing to swap her bundle should she get the opportunity to do so.
This indicates the individual's perceived envy (but not the degree of envy).
Participants’ responses are aggregated into a single \emph{swap rate}, the percentage of scenarios where a swap was chosen, measuring aggregate perceived fairness under each treatment.

\paragraph{\textbf{Epistemic vs. Counterfactual Envy.}}
Our results show that \HEF{k} allocations are perceived to be fairer than in the \sEF{1} and \EF{1} treatments.
In particular, we show that there is a statistically significant difference between swap rates     of \HEF{k} and both \sEF{1} and \EF{1} treatments ($p<0.001$). Participants under the \HEF{k} treatment displayed the lowest swap rate, followed by \sEF{1} and then \EF{1} (Section~\ref{sec:perceived}). We subsequently control for the effect of variables such as instance size, allocation balance (defined in Section \ref{sec:data_set}), and scenarios for which it is optimal to swap, and find that the qualitative results still hold.

Additionally, we study \emph{cognitive effort}, as measured by response time and self-reports of scenario difficulty, to understand how treatment affects participants' reasoning.
% and how it correlates with perceived fairness.
We find that there is a significant difference in the cognitive effort exercised by participants, measured by response time and self-reports of difficulty, between the \HEF{k} and both \sEF{1} and \EF{1} treatments ($p<0.001$) (Section~\ref{sec:cognitive}). 
Hence, perceived fairness appears correlated with the cost of increased task complexity.

\paragraph{\textbf{Human Subject Dataset.}}
% \HH{first of its kind to the best of our knowledge; we plan to make it freely available; it contains 166 questions involving X and Y.}
To conduct our analysis we generated a novel data set of 166 scenarios, each consisting of a fair division instance, allocation that satisfies one of the investigated fairness properties, and anonymized choices made by participants. The number of scenarios satisfying each instance size, allocation balance, and allocation fairness property may be found in \cref{tab:survey_question_properties} in the appendix. This data set is the first of its kind, to the best of our knowledge, and will be made publicly available upon publication.

\subsection{Related Work}
\label{sec:related_work}

Our work is in line with research empirically validating fairness notions and theories of distributional preferences. 
While it is evident people 
trade off self-interest for fairness \citep{kahneman1986fairness}, it is still not clear to what extent and which theories of fairness are the most valid. Prior experiments have employed several methods to evaluate perceived fairness of allocations, often asking participants which they prefer. For instance,
\citet{herreiner2009envy} empirically investigated \EF{} in a free-form bargaining experiment. In their setting, participants had subjective preferences over goods and collaborated with another participant to choose the allocation (see also \citet{herreiner2010inequality}). The authors subsequently analyzed the fairness and efficiency of the chosen allocations. 
This work is most similar to ours, except that we measure the envy experienced by participants and focus on the relative fairness of relaxations of \EF{}.

\citet{herreiner2009envy}'s work follows a tradition of questionnaire methodology for evaluating distributive justice, popularized by
\citet{yaari1984dividing} and \citet{konow2003fairest},
% used questionnaires to
who asked whether participants perceive given allocations as just or not (see also \citet[Chapter 9]{gaertner2009primer}). \citet{herreiner2007distributing} also asked participants to choose which of a set of allocations was the most fair. 
While these studies provide some evidence in favor of certain fairness notions,
payoffs were identical, so
intrapersonal theories like \EF{} could not be tested. 
In this vein, 
\citet{engelmann2004inequality} ran an experiment were participants would, with some probability, received the allocation of money they chose.
Their aim was to compare the explanatory power of   distributional preferences models by \citet{fehr1999theory},  \citet{bolton2000erc}, and \citet{charness2002understanding}.
% \footnote{See also the subsequent experiments by \citet{kritikos2001distributional} and \citet{bereby2005fairness} and the back-and-forth discussion of \citet{bolton2006measuring}, \citet{fehr2006inequality}, and \citet{engelmann2006inequality}.} 

A separate line of work by Lee and colleagues focused on perceived fairness of algorithmic decision-making.
Participants in \citet{Lee17:Mediation}'s study perceived allocations prescribed by Spliddit\footnote{\url{http://www.spliddit.org/}} to be less fair
than those chosen in group discussions one third of the time.
The authors explain this distinction as the algorithms excluding the effects of individual participation, interpersonal power, and altruism on fairness.
\citet{lee18:understanding} suggested that perceived fairness depends on task characteristic, which helps motivate our current study on cognitive effort. Lee's participants recognized that algorithms produce less fair decisions on tasks requiring human skills, such as those requiring subjective judgement, but equally fair on mechanical tasks, such as processing data.
\citet{Lee19:Justice} measured the effect of transparency and outcome control (i.e., the ability to manually adjust prescribed outcomes) on perceived fairness of \EF{1} allocations prescribed by Spliddit.
They showed that perceived fairness increased after participants were given an opportunity to modify the allocation, either individually or through group discussions.
These studies substantially differ from ours in that there is an impact of personal image and social pressure in bargaining and collective decision-making, which may provide a justification for inequality aversion. Furthermore, there is a sense of agency within discussions or ability to modify the outcome, which may result in higher satisfaction via the \textit{IKEA Effect}.\footnote{The IKEA effect is a cognitive bias in which people tend to value on products they helped to create highly~\citep{norton2012ikea}.}

Other empirical research includes \citet{kyropoulou22:fair}, who tested the effect of participants' strategic behavior in choosing allocations of divisible resources on total envy.
Separately, \citet{konig2019fair} measured the suitability of two well-adopted matching mechanisms, the Boston mechanism and assortative matching, under the \textit{veil of ignorance} \citep{rawls2004theory} assumption. They concluded that which procedure participants prefer depends on how much autonomy they have to report their preferences.
The empirical validity of fairness axioms in cooperative games \citep{d2020testing,Clippel22:Cooperative} and machine learning \citep{Chakraborti2022contrastive} has also been studied.

\section{Model and Solution Concepts}
\label{sec:model_defs}

\begin{paragraph}{Model.} For any $k\in \bn$, we define $[k]:=\{1,\dots,k\}$. An instance of the fair division problem is a tuple $\mi=\langle N, M, V \rangle$, where $N\coloneqq [n]$ is a set of $n$ {\em agents}, $M\coloneqq [m]$ is a set of $m$ {\em goods}, and $V\coloneqq \{v_1,\dots,v_n\}$ is a {\em valuation profile} that specifies for each agent $i\in N$ her preferences over the set of all possible {\em bundles} $2^M$. This {\em valuation function} $v_i:2^M\to\bn\cup\{0\}$ maps each bundle to a non-negative integer. 
We write $v_{i,j}$ instead of $v_{i}(\{j\})$ for a single good $j\in M$. 
We assume that the valuation functions are {\em additive} so that for any $i\in N$ and $S\subseteq M$, $v_i(S)\coloneqq\sum_{j\in S}v_{i,j}$, where $v_i(\emptyset)=0$.
\end{paragraph}

\begin{paragraph}{Allocation.} An allocation $A\coloneqq(A_1,\dots,A_n)$ is a (complete) $n$-partition of the set of goods $M$, where $A_i\subseteq M$ is the bundle allocated to agent $i\in N$. 
\end{paragraph}

\begin{dfn}[Envy-freeness] An allocation $A$ is:
\begin{enumerate*}[label=(\roman*)]
    \item {\em envy-free} (\EF{}) if for every pair of agents $h,i\in N$, $v_i(A_i)\ge v_i(A_h)$~\citep{foley1966resource},
    \item {\em strongly envy-free up to one good} (\sEF{1}) if for each agent $h\in N$ such that $A_h\neq\emptyset$, there exists a good $g_h\in A_h$ such that for every $i\in N$, $v_i(A_i)\ge v_i(A_h \backslash \{g_h\})$~\citep{conitzer19:group}, and 
    \item {\em envy-free up to one good} (\EF{1}) if for each pair of agents $h,i\in N$, there exists a good $g_h\in A_h$ such that $v_i(A_i)\ge v_i(A_h \backslash \{g_h\})$~\citep{Lipton04:Approximately,B11combinatorial}.
\end{enumerate*}
\label{dfn:envy-free}
\end{dfn}

\begin{dfn}[Envy-freeness with hidden goods] An allocation $A$ is {\em envy-free up to $k$ hidden goods} (\HEF{k}) if $\exists S\subseteq M$, $|S|\le k$, such that for every pair of agents $h,i\in N$, we have that $v_i(A_i)\ge v_i(A_h \backslash S)$~\citep{Hosseini2020:Fair}.
\label{dfn:hefk}
\end{dfn}

By the above definitions, \EF{} implies \sEF{1}, which implies \EF{1} and subsequently \HEF{k} for some $k \leq m$. Moreover, an allocation is \EF{} if and only if it is \HEF{0} and $\forall k \geq 0$ \HEF{k} implies \HEF{(k+1)} \citep{Hosseini2020:Fair}. 
To disambiguate these classes and reduce confusion, we impose the following technical qualifications throughout this paper.
% We distinguish these classes throughout this paper with the following qualifications. 
First, we recognize two variants of envy-freeness up to one good by discerning allocations that are \EF{1} but not \sEF{1}. Through an abuse of notation, we henceforth label this \emph{weak} variant ``\EF{1}.'' Both variants (weak and strong) correspond to the \emph{counterfactual} removal of goods when agents have full information about the entire allocation. 
Second, 
for any \HEF{k} allocation with hidden set $S$, 
each agent $i$ knows their own bundle $A_i$ but only has partial information about the goods in the bundle of any other agent $h$.
Then, $i$ has no envy among the observable (partial) allocation (i.e., $v_i(A_i) \geq v_i(A_h \backslash S)$). 
Furthermore, we assert that $|S| = k$ and that $A$ is not \HEF{k'} with respect to any strict subset $S' \subset S$, where $|S'|=k'<k$.

\begin{figure*}[t] %[ht] (4/18/23)
\footnotesize
    \begin{tabular}{ccc}
         \begin{tabular}{p{0.5em}|cccccc}
            & $g_1$ & $g_2$ & $g_3$ & $g_4$ & $g_5$ & $g_6$ \\
            \hline
            $v_1$ & \cellcolor{blue!25}$2$ & \cellcolor{blue!25}\sqboxed{\diamondboxedcompact{$2$}} & $4$ & $1$ & $1$ & $4$ \\
            $v_2$ & $1$ & $4$ & \cellcolor{blue!25}\circledcompact{\diamondboxedcompact{$1$}} & \cellcolor{blue!25}$1$ & $4$ & $1$ \\
            $v_3$ & $4$ & $1$ & $3$ & $3$ & \cellcolor{blue!25}\sqboxed{\circled{$2$}} & \cellcolor{blue!25}$2$ \\
        \end{tabular}
        &
        \begin{tabular}{p{0.5em}|cccccc}
            & $g_1$ & $g_2$ & $g_3$ & $g_4$ & $g_5$ & $g_6$ \\
            \hline
            $v_1$ & \cellcolor{blue!25}\diamondboxed{$2$} & $2$ & $4$ & $1$ & \cellcolor{blue!25}\sqboxed{$1$} & $4$ \\
            $v_2$ & $1$ & $4$ & \cellcolor{blue!25}\circled{$1$} & \cellcolor{blue!25}$1$ & $4$ & $1$ \\
            $v_3$ & $4$ & \cellcolor{blue!25}\sqboxed{$1$} & $3$ & $3$ & $2$ & \cellcolor{blue!25}\circled{$2$} \\
        \end{tabular}
         & 
         \begin{tabular}{p{0.5em}|cccccc}
            & $g_1$ & $g_2$ & $g_3$ & $g_4$ & $g_5$ & $g_6$ \\
            \hline
            $v_1$ & $2$ & $2$ & $4$ & \cellcolor{blue!25}$1$ & $1$ & \cellcolor{blue!25}$4$ \\
            $v_2$ & $1$ & \cellcolor{blue!25}$4$ & $1$ & $1$ & \cellcolor{blue!25}$4$ & $1$ \\
            $v_3$ & \cellcolor{blue!25}$4$ & $1$ & \cellcolor{blue!25}\circled{$3$}
            & $3$ & $2$ & $2$ \\
        \end{tabular}
        \\
        (a) \sEF{1} & (b) \EF{1} & (c) \HEF{k}
    \end{tabular}
    \centering
    \caption{Allocations satisfying (a) \sEF{1}, (b) \EF{1} and (c) \HEF{k} for a fair division problem instance. Elements marked by a circle, rectangle, and diamond must be hidden or counterfactually removed to eliminate the envy from agents $1$, $2$, and $3$ respectively.
    }
    \label{fig:model_example}
\end{figure*}

\begin{examplex}[Epistemic (\HEF{}) vs. Counterfactual (\sEF{1} and \EF{1})]
\label{eg:sEF1_EF1}
\cref{fig:model_example} demonstrates three allocations for the same instance with three agents $1,2,3$ and six goods $g_1, \dots, g_6$. These are demonstrated by the underlined elements in subfigures (a), (b), and (c), satisfying \sEF{1}, \EF{1}, and \HEF{1} respectively.
Elements outlined by a circle, rectangle, and diamond must be counterfactually removed (for \sEF{1} and \EF{1}) or hidden (for \HEF{1}) to eliminate the envy of agents $1$, $2$, and $3$ respectively. 

Consider the \EF{1} allocation $A$ where $A_1 = \{g_1, g_5\}$, $A_2 = \{g_3, g_4\}$, and $A_3 = \{g_2, g_6\}$. Although agent $1$ is envious of agents $2$ and $3$, we have $v_1(A_1) \geq v_1(A_2 \backslash \{g_3\})$ and $v_1(A_1) \geq v_1(A_3 \backslash \{g_6\})$. For the \HEF{1} allocation, rather, agent $1$ is not envious of agent $3$ because they only observe a partial allocation: $v_1(A_1) \geq v_1(A_3 \backslash S)$ where $S = \{g_3\}$. Agent $3$ is not envious of agent $1$ because they observe the entire allocation and $v_3(A_3) \geq v_3(A_1)$.

Notice that at most a single good is outlined in each agent's bundle in the \sEF{1} allocation, whereas multiple goods may be outlined in each bundle in the \EF{1} allocation.
\end{examplex}

\section{Experimental Design}

\begin{figure*}
\centering
\begin{tabular}{c|c}
    \includegraphics[width=0.47\textwidth]{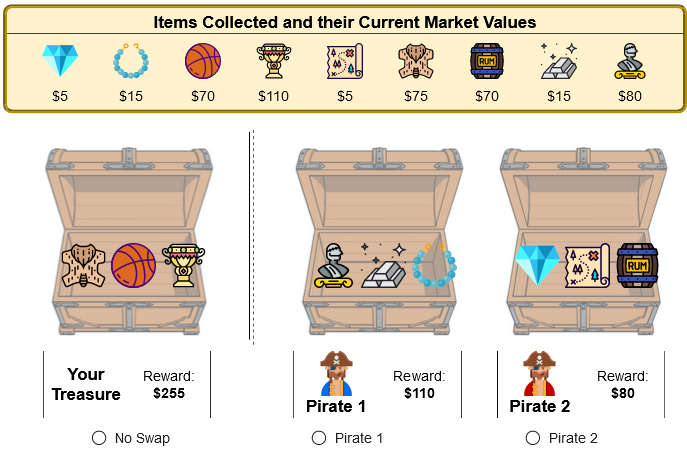}
     & 
    \includegraphics[width=0.47\textwidth]{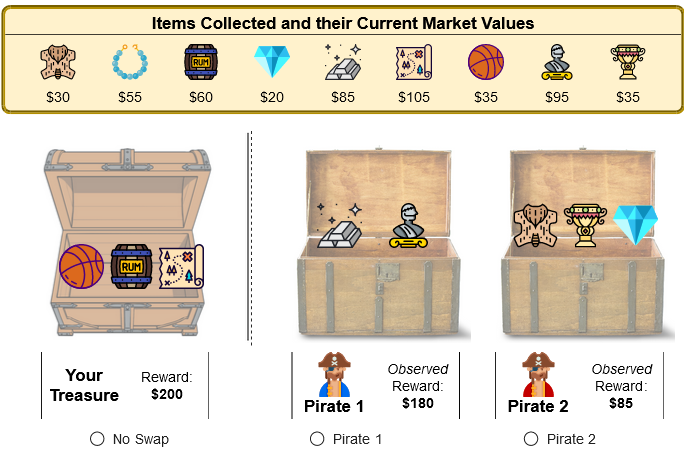}
     \\
    (a) & (b)
\end{tabular}
\caption{ Sample scenarios from the (a) \EF{1} and \sEF{1} treatments and (b) \HEF{k} treatment. 
In the \EF{1} and \sEF{1} treatments, participants have a `birds-eye view' of all goods in all bundles.
In the \HEF{k} treatment, participants observe only the revealed goods from other pirates' `upright' boxes.
All \HEF{k} treatment participants complete a training tutorial emphasizing this point.}
\label{fig:question}
\end{figure*}

We conducted an empirical study to compare the perceived fairness of 
multiple
relaxations of envy-freeness---\sEF{1}, \EF{1}, and \HEF{k}---using a gamified pirate scenario (see \cref{fig:question}).
Participants were split into three treatments
and given twelve scenarios.
In each scenario, the participant was assigned the role
of one member of a crew of pirates (agents) whose captain (a central authority) wished to divide goods, the spoils of a recent adventure, among the crew. Each scenario consisted of a number of goods, presented in a \emph{marketplace}, and the bundles of (revealed) goods for each pirate in an allocation determined by the captain. Participants' \emph{subjective} values for each bundle were determined by the given instance and the perspective of the participant. For instance, a participant could be offered the instance and allocation demonstrated by  \cref{fig:model_example}(a) from the perspective of agent $1$ and would value their bundle at $v_1(A_1) = 1+1 = 2$. Alternatively, their value for $A_1$ from the perspective of agent $3$ would be $v_3(A_1) = 4+1 = 5$.

Given this information, participants were 
asked whether they wanted to {\emph swap} their bundle with that of another pirate of their choice, in its entirety, or keep their initial bundle. 
Participants had a stake in the outcome of their choices: they received
a bonus payment if the total value of goods they collected surpassed a threshold. Therefore, choosing to swap bundles indicates the participant's envy and perceived unfairness.
We measured participants’ \emph{swap rate}, the percentage of scenarios where a swap was chosen,
and compared treatments
using the Chi-square ($\Chi^2$) \citep{mchugh2013chi} and Fisher's exact tests \citep{kim2017statistical}. We then compared treatments upon segmenting our data by (i) the number of agents and goods (instance size), (ii) the distribution of goods across agents (allocation balance), and (iii) whether it is optimal for participants to swap or not, including the value of hidden goods (optimal choice).

\begin{paragraph}{Treatment details}

Each participant was subjected to exactly one of three treatments -- \sEF{1}, \EF{1}, and \HEF{k} -- 
corresponding with the fairness property satisfied by their allocations.
% such that each allocation presented to the participant satisfied the same relaxation of \EF{}.
Across treatments, participants were shown their subjective values of the visible portions of the bundles of each agent. 
Participants in the \sEF{1} and \EF{1} treatments had full information about the allocations (see \cref{fig:question}(a)). 
Participants assigned the \HEF{k} treatment were shown their own bundles but only the visible portions of other agents' bundles (recall Definition \ref{dfn:hefk}; see \cref{fig:question}(b)). 
We explained through a tutorial that the visible allocation was incomplete by detailing the possibilities of the missing information: some goods may be allocated to and hidden by other pirates or discarded altogether. 
Participants could therefore enumerate the possible values of the other agents' bundles. 

Our study employed 120 mutually exclusive participants for each of three Human Intelligence Tasks (HITs), corresponding to the three treatments, in Amazon's Mechanical Turk platform, totaling 360 participants. 
Our study was single-blind; participants were not aware of their treatment.
\end{paragraph}

\begin{paragraph}{Perceived fairness.}
We measured perceptions of one aspect of fairness, envy, via swaps.
Specifically, given an allocation $A$, we say that agent $i$ \emph{swaps} her bundle $A_i$ with agent $h$ if the agents exchange all goods within their bundles (including hidden goods). 
An agent choosing to swap bundles indicates that they are envious of another agent and thus does not perceive their bundle $A_i$ as fair. 
We call the proportion of participants that swap under $A$ its empirical \emph{swap rate}, representing the aggregate perceived fairness of the scenario.
\end{paragraph}

\begin{paragraph}{Incentives.}
In order to realize the assigned in-game valuations as real-world \emph{value}, participants were incentivized to accumulate high-value bundles throughout the survey. Specifically, each participant was eligible to receive two payments: 
\begin{enumerate*}[label=(\arabic*)]
    \item a \emph{base} payment of \$0.50
    for completing the survey in its entirety, and
    \item a \emph{bonus} payment of \$0.50
    for accumulating at least \$2000 worth of goods through all scenarios as measured by participants' assigned subjective valuations.
\end{enumerate*} 
Hence, we are able to emulate a real-world setting through our experiment with fictional pirate-related goods.

Note that within the \HEF{k} treatment, participants accumulate the values of any hidden goods of their chosen bundle as well.
The bonus threshold was also chosen to encourage participants to pay greater attention to the study and not choose randomly for each scenario.
We determined the threshold by computing the minimum and maximum total value any participant could obtain on any survey using our data set. We then chose \$2000 which falls between between 71\% and 84\% for these ranges. 
\end{paragraph}

\begin{paragraph}{Response qualifications.}
In order to obtain high quality responses, participation in our study was restricted to Mechanical Turk workers who 
\begin{enumerate*}[label=(\alph*)]
    \item had at least an 80\% approval rate on previous tasks, 
    \item had completed at least 100 tasks, 
    \item were located in either the United States or Canada\footnote{We restricted location to ensure language proficiency and prevent any potential issues due to linguistic barriers.},
    \item had a Master's qualification\footnote{Workers with Master's qualification, determined by Mechanical Turk, are those who ``have consistently demonstrated a high degree of success in performing a wide range of HITs across a large number of Requesters.'' See \url{https://www.mturk.com/worker/help}.} on the Mechanical Turk platform, and 
    \item had not attempted or taken the survey before.
\end{enumerate*}
Through the experiment we adjusted the minimum HIT approval rate (\%) and minimum number of HITs approved that were necessary in order to attract Mechanical Turk Workers to participate; see \cref{tab:response_qualifications} in the appendix.
\end{paragraph}

\subsection{Data Set}
\label{sec:data_set}

The scenarios were sampled from a novel data set of 
166
scenarios, each consisting
of a fair division instance, an allocation partitioning the goods, 
and an assignment of the participant to one of the agent's perspectives. 
% \footnote{
% Data will be made publicly available online upon publication.
% }

\paragraph{Instances.} We generated twenty-eight instances involving nine or ten goods: twenty-one \smallq{} instances with three agents and seven \largeq{} instances with five agents. Each valuation $v_{i,j}$ for $i \in N$ and $j \in M$ was
sampled uniformly at random from $\{5,10,\dots,120\}$ for \smallq{} instances and $\{0,10,\dots,150\}$ for \largeq{} instances.

\begin{paragraph}{Allocations.}
For each instance, we computed three allocations
satisfying
\sEF{1}, \EF{1}, and \HEF{k}
for a pre-specified $k \in \{0,1,2\}$ for the corresponding treatments. 
Allocations were computed by randomly shuffling goods across agents until the desired properties were achieved. As we observe in Example \ref{eg:sEF1_EF1}, \EF{1} allocations can sometimes require the counterfactual removal of a larger number of goods than \sEF{1} allocations. To reflect this, and emphasize the distinction between \EF{1} and \sEF{1} allocations in our experiments, we picked \EF{1} allocations that require at least $n+2$ goods to be counterfactually removed to eliminate envy among agents.

There were two levels of \emph{balance} for allocations. A \bal{} allocation gives every agent a bundle of equal size, three (respectively, two) goods to each agent in a small (respectively, large) instance. In an \unbal{} allocation, agents may have bundles consisting of different number of goods, with bundle sizes $(2,4,4)$ for \smallq{} instances and either $(4,2,2,1,1)$ or $(3,2,2,2,1)$ for \largeq{} instances. 
\end{paragraph}

\begin{paragraph}{Perspective.}
Participants were randomly assigned to assume the role of either the first or last (i.e., third or fifth) agent in the instance. 
Providing two perspectives expanded our data set and enabled participants to have different goods in their bundles for the same instances. This did not bias our results as valuations were randomly generated and allocations did not depend on agents' identities. 
\end{paragraph}

\begin{paragraph}{Scenario properties.}

\begin{figure}[t]
  \centering 
  \includegraphics[width=.45\linewidth]{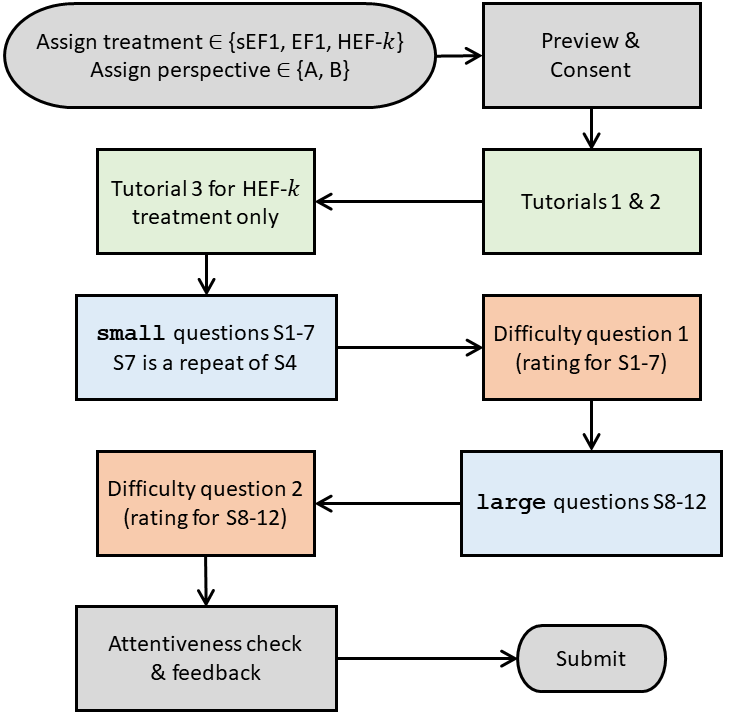}
  \caption{The workflow of a participant.
  }
  \label{fig:turker_workflow}
\end{figure}

Our $166$ scenarios were made with the following combinations: 
$63$ allocations were affiliated with \smallq{} instances, of which $45$ were \bal{} and $18$ were \unbal{}, while $20$ allocations were affiliated with \largeq{} instances, of which $14$ were \bal{} and $6$ were \unbal{}.
\cref{tab:survey_question_properties} in the appendix presents the number of allocations in each treatment succinctly.
Each of the these $83$ allocations provided two scenarios to the data set, corresponding to two perspectives we offered participants, yielding the $166$ total scenarios.
\end{paragraph}

\subsection{Survey Outline}
\label{sec:survey_outline}

Participants undertook the following workflow (see \cref{fig:turker_workflow}).
First, participants gave their consent to partake in our IRB-approved study after being informed of the study description, benefits, risks, rights, and project manager contact information.
After being assigned a treatment and a randomly determined perspective, they subsequently answered twelve scenario questions, two questions soliciting scenario difficulty, and two attentiveness check questions. The scenarios were organized into four sections, each consisting of scenarios of different instance size and allocation balance that were selected uniformly at random from the appropriate data set, and then randomly permuted within the section. A complete survey
therefore consisted of:
\begin{itemize}[leftmargin=*]
    % \item Tutorial questions.
    \item Section 1 (S1--3): 3 \sbal{} scenarios. If the treatment is \HEF{k}, then $k\in\{0,1,2\}$ respectively.
    \item Section 2 (S4--7): 3 \sunbal{} scenarios     followed by S7 which is a repeat of S4. If the treatment is \HEF{k}, then $k\in\{0,1,2\}$ respectively for (S4--S6).
    \item Difficulty: self-reported rating for \smallq{} scenarios.
    \item Section 3 (S8--10): 3 \lbal{} scenarios. If the treatment is \HEF{k}, then $k\in\{0,1,2\}$ respectively.
    \item Section 4 (S11-12): 2 \lunbal{} scenarios. If the treatment is \HEF{k}, then $k=1$.
    \item Difficulty: self-reported rating for \largeq{} scenarios.
    % \item Final page with two attentiveness check questions.
\end{itemize}

\begin{paragraph}{Tutorials.}
All participants were required to correctly answer a few tutorial questions prior to the scenarios.

The first tutorial taught participants that the value of a bundle was equal to the sum of values of the goods inside that bundle. Participants were presented with a bundle consisting of three goods, which were highlighted in the marketplace, and were asked to compute the bundle's value.

The second tutorial taught participants that whether they received a monetary bonus upon completing the survey is dependent on the total value they collect throughout its course. The participants were presented with three bundles, similar to Figure \ref{fig:question}(a), and were asked if they wanted to keep their bundle (left) or swap it with either Pirate 1's bundle (middle) or Pirate 2's bundle (right). The bundle with the highest value was enforced as the correct choice. 

\HEF{k} treatment participants were provided a third tutorial designed to teach them about goods in the marketplace that were not visibly allocated. Participants were presented with three bundles, similar to Figure \ref{fig:question}(b), and were told that the missing goods 
may be either allocated to and hidden by the other pirates or discarded altogether. 
Participants were asked about the maximum number of goods that could be found in any one pirate's bundle, thus requiring them to reason about the location of  missing goods.
\end{paragraph}

\begin{paragraph}{Self-reported difficulty.}
The groups of seven \smallq{} and five \largeq{} scenarios were each succeeded by a question asking participants to rate the difficulty of the scenarios on a 5-point Likert scale from Very Easy (1) to Very Hard (5).
\end{paragraph}

\begin{paragraph}{Attentiveness check questions.}
We incorporated many checks to ensure high quality responses from attentive human participants and dissuade fraud, which is a known problem for Mechanical Turk \citep{kennedy2020shape}. Prior to the tutorial, participants answered a simple arithmetic problem to ensure they were not bots. On the final page, they answered (1) their favorite good and (2) final comments or questions. We presumed that we could identify inattentive participants giving poor quality data, as they would not be able to answer these prompts appropriately. We did not find any participants' responses to be of poor quality by these measures, so we did not discard any responses. 
% Additional details about participant qualifications can be found in \cref{apx:additional_details}.
\end{paragraph}

\section{Experimental Results}
\label{sec:exp_results}

\begin{table*}[t]
    \footnotesize{}
    % \small
    \centering
    \caption{ $p$-values of the test statistic for testing the independence of swap rates and treatments under different pairs of treatments and adjusting for different variables. The $\Chi^2$ test is used except when the $p$-value is annotated with a ``$\dagger$'', in which it is the result of the Fisher's exact test. ``p: \ns{}'' denotes non-significance.
    % The $p$-value of the test statistic is represented as follows: a cell labeled \ns{} (not significant) implies that $p >0.05$, $\medstar$ for $p \in (0.01, 0.05]$, $\medstar \medstar$ for $p \in (0.001, 0.01]$, and $\medstar \medstar \medstar$ for $p < 0.001$. 
    }
    \label{tab:perceived_fairness}
    \begin{tabular}{|c|c||c|c|c||c|c|c|}\hline
        \multirow{2}{*}{\bf Variable} & \multirow{2}{*}{\bf value} & \multicolumn{6}{c|}{\bf Pairs of Treatments} \\ \cline{3-8}
         & & \HEF{k}, \sEF{1} & \HEF{k}, \EF{1} & \sEF{1}, \EF{1} & \HEF{0}, \sEF{1} & \HEF{1}, \sEF{1} & \HEF{2}, \sEF{1} \\ \hline \hline
        %%%%%%%%%%%%%%%%%%%%%%%%%%%%%%%%%%%%%%
        \multicolumn{2}{|c||}{All scenarios} & 
        $p < 0.001$ & 
        $p < 0.001$ & 
        $p < 0.001$ & 
        $p < 0.001$ & 
        $p < 0.001$ & 
        $p < 0.001$ \\
        \hline
        %%%%%%%%%%%%%%%%%%%%%%%%%%%%%%%%%%%%%%%%%%%%
        \multirow{2}{*}{\makecell{Optimal\\ Choice}}
        & \stayopt{} & 
        $p < 0.001$ & 
        $p: \ns{}$ & % $p=0.15$
        $\dagger p < 0.01$ & 
        $p < 0.001$ & 
        $p < 0.001$ & 
        $p:\ns{}$ \\
        \cline{2-8}
        & \swapopt & 
        $p < 0.001$ & 
        $p < 0.001$ & 
        $p < 0.05$ & 
        \makecell{\texttt{N/A}} & 
        $p < 0.001$ & 
        $p < 0.001$ \\
        \hline
        %%%%%%%%%%%%%%%%%%%%%%%%%%%%%%%%%%%%%%%%%%%
        \multirow{2}{*}{\makecell{Instance\\ Size}} 
        & \smallq{} &
        $p < 0.001$ & 
        $p < 0.001$ & 
        $p < 0.001$ & 
        $p < 0.001$ & 
        $p < 0.001$ & 
        $p < 0.001$ \\
        \cline{2-8}
        & \largeq{} &
        $p < 0.001$ & 
        $p < 0.001$ & 
        $p < 0.001$ & 
        $p < 0.001$ & 
        $p < 0.001$ & 
        $p < 0.001$ \\
        \hline
        %%%%%%%%%%%%%%%%%%%%%%%%
        \multirow{2}{*}{Balance}
        & \bal{} &
        $p < 0.001$ & 
        $p < 0.001$ & 
        $p < 0.001$ & 
        $p < 0.001$ & 
        $p < 0.001$ & 
        $p < 0.001$ \\
        \cline{2-8}
        & \unbal{} & 
        $p < 0.001$ & 
        $p < 0.001$ & 
        $p < 0.001$ & 
        $p < 0.001$ & 
        $p < 0.001$ & 
        $p < 0.001$ \\
        \hline
        %%%%%%%%%%%%%%%%%%%%%%%%%%%%%%%%%%%%%%%%
        \multicolumn{2}{|c||}{Repeated scenario (S7)} & 
        $p < 0.001$ & 
        $p < 0.001$ & 
        $p < 0.001$ & 
        $p < 0.001$ & 
        $p < 0.01$ & 
        $p < 0.001$ \\
        \hline
    \end{tabular}
\end{table*}

We test the empirical swap rate of each treatment as a measure for perceived fairness across all scenarios and while controlling for several variables. We further partition the \HEF{k} treatment into sub-treatments---\HEF{0}, \HEF{1}, and \HEF{2}---and compare their swap rates with \sEF{1}, with a focus on whether increasing the number of hidden goods affects perceived fairness. Separately, we compare the effect of treatment and size of instance on participants' cognitive effort, as measured by response time and self-reports of difficulty, for answering the scenarios.

We are particularly interested in 
whether swap rates differ between treatments when a participant's \emph{optimal} (i.e., value-maximizing) choice is to either \emph{stay} or \emph{swap} bundles. This is because participants may be biased to accept their default bundle and maintain the status quo rather than make adjustments \citep{samuelson1988status}. Moreover, \HEF{k} differs from the other treatments in that participants may not have enough information to distinguish which bundle is optimal, despite it being apparently optimal. Our work is the first to study whether perceived fairness, as measured by swap rates, differ depending on optimal choice. 
% This raises the  question of whether 

\subsection{Perceived Fairness} \label{sec:perceived}

We formalize our research questions as follows:

\begin{paragraph}{Research Questions:}

For any two treatments $X,Y \in \{\sEF{1},\EF{1},$ $\HEF{k}\}$ or $\{\sEF{1}, \HEF{0}, \HEF{1}, \HEF{2}\}$,
do swap rates differ between $X$ and $Y$
overall and when adjusted independently for the variables:
\begin{enumerate*}[label=(\roman*)]
    \item {\em instance size}: \smallq{} or \largeq{},
    \item {\em allocation balance}: \bal{} or \unbal{}, and 
    \item {\em optimal choice}: whether the value-maximizing choice is to keep the participant's initial bundle (\stayopt{}) or to swap bundles (\swapopt{})?
\end{enumerate*}

{\em Null Hypothesis:} Swap rate is independent of treatment.

{\em Alternate Hypothesis:} Swap rate depends on treatment.

\end{paragraph}

\begin{figure}[t]
  \centering %[htp]
    \includegraphics[width=\linewidth]{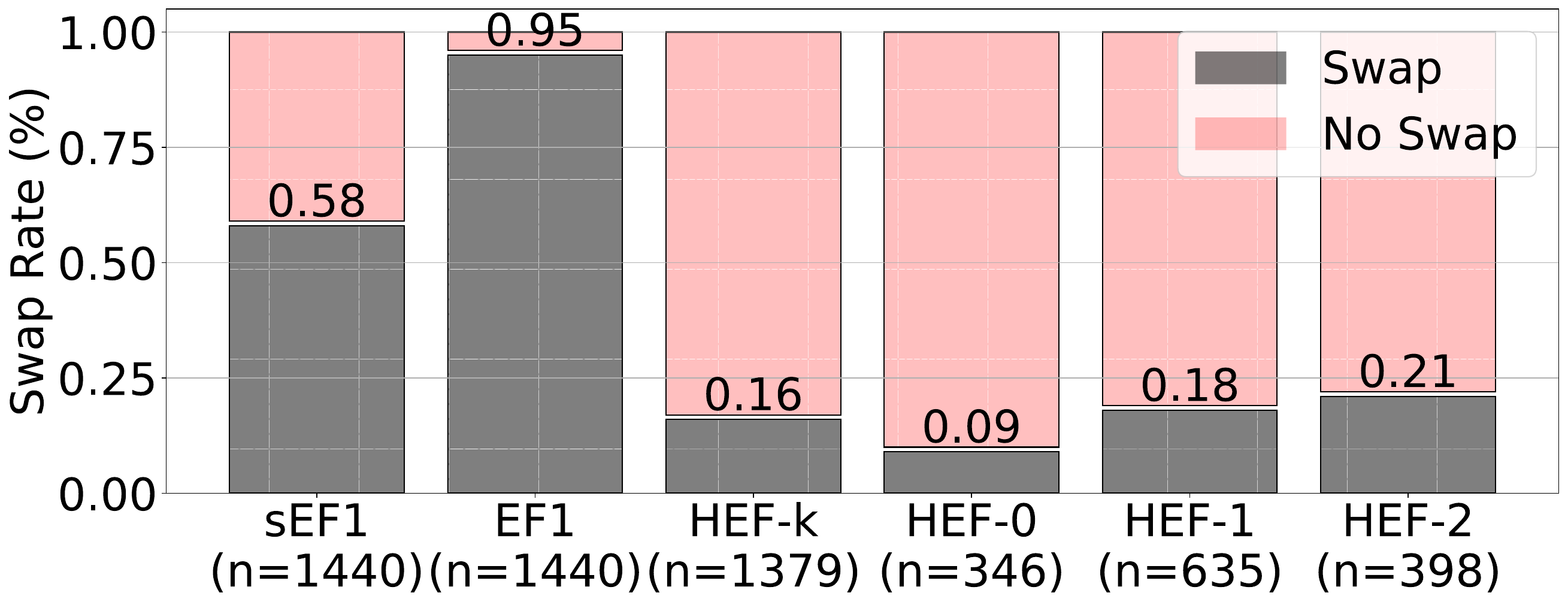}
  \caption{Swap rates per treatment, all scenarios. Here, $n$ is the number of scenarios per treatment.}
  \label{fig:mosaic-all_qs}
\end{figure}

Our experiments provide statistically significant evidence for rejecting the null hypothesis that swap rate is independent of treatment. We draw this conclusion using the Chi-square ($\Chi^2$) test with $p < 0.05$ for all combinations of pairs of treatments and values for the different confounding variables in our study,
% and $p < 0.05$ for the remaining combinations.

Table~\ref{tab:perceived_fairness} summarizes our findings. 
% For each pair of treatments $X$ and $Y$, identified by the column `$X, Y$', and for certain confounding variables, we present the ratio of swap rates between $X$ and $Y$. For example, the ratio of swap rates from all \HEF{k} scenarios to all \sEF{1} scenarios is $0.286$.
In the appendix we present Tables 
\ref{tab:perceived_fairness_full} and 
\ref{tab:perceived_fairness_cramer_v} which includes more specific information about the 
$p$-values of the $\Chi^2$ and Fisher's exact test statistics and 
\emph{effect size}, as measured by Cramer's V \citep{kim2017statistical}, about the tests. 
% We include further tests conditioning on instance size and allocation balance.
% 
% 
% 
Our main finding is that 
% across all scenarios, 
\begin{enumerate*}[label=(\arabic*)]
    \item the perceived envy of \HEF{k} is significantly lower than that of either \sEF{1} and \EF{1}, and 
    \item \sEF{1} allocations are less likely to be perceived as unfair than \EF{1} allocations
\end{enumerate*}, as we show in \cref{fig:mosaic-all_qs}. This holds true upon adjusting for instance size (\smallq{} or \largeq{}) and the allocation balance (\bal{} or \unbal{}), and among scenarios where \swapopt{}. Thus, our main takeaway message is:

\begin{center}
{\em Allocations that are visibly envy-free through hiding goods are perceived to be fairer than allocations that are counterfactually envy-free 
via removing goods.
}
\end{center}

\begin{paragraph}{Segmented Data.}
Upon realizing this conclusion, we segment our data 
to draw additional insights. In particular, among \HEF{k} allocations, swap rate increases as the number of hidden goods increases (\cref{tab:hef_k_perceived_fairness} in the appendix): \HEF{0} allocations induce less envy among the participants than either \HEF{1} or \HEF{2}. This is perhaps because as more goods are hidden, participants are more cautious, more uncertain about the allocation's fairness, and spend more time on average to choose bundles (see \cref{fig:plot-times-2g} in the appendix). Further studies may be necessary to explain these results.

% Additionally, to control for any preferential bias toward our choices of pirate-related goods, we repeated a scenario and replaced the goods with identically-shaped gems of different colors (see ``Repeated scenario (S7)'' in Table \ref{tab:perceived_fairness}). This test is described further in the appendix.
\end{paragraph}

\begin{table*}[t] %[htp]
    % \tiny
    % \footnotesize{}
    \small
    \centering
    \caption{ $p$-values of the test statistic for testing the independence of swap rates and optimal choice under different treatments. The $\Chi^2$ test is used except when the $p$-value is annotated with a ``$\dagger$'', in which it is the result of the Fisher's exact test. 
    % The $p$-value of the test statistic is represented as follows: a cell labeled \ns{} (not significant) implies that $p >0.05$, $\medstar$ for $p \in (0.01, 0.05]$), $\medstar \medstar$ for $p \in (0.001, 0.01]$, and $\medstar \medstar \medstar$ for $p < 0.001$.
    }
    \label{tab:stay_vs_swap_is_opt}
    \begin{tabular}{|c||c|c|c|c|c|c|}\hline
        \textbf{Treatment} & \sEF{1} & \HEF{k} & \EF{1} & \HEF{0} & \HEF{1} & \HEF{2} \\
         \hline
        \makecell{\swapopt{}\\/ \stayopt{}} & 
        $p < 0.001$ & 
        $p < 0.001$ & $\dagger p < 0.001$ & 
        \texttt{N/A} & 
        $p < 0.001$ & 
        $p < 0.001$ \\ 
         \hline
    \end{tabular}
\end{table*}

\begin{figure}[t]
  \centering %[htp]
    \includegraphics[width=\linewidth]{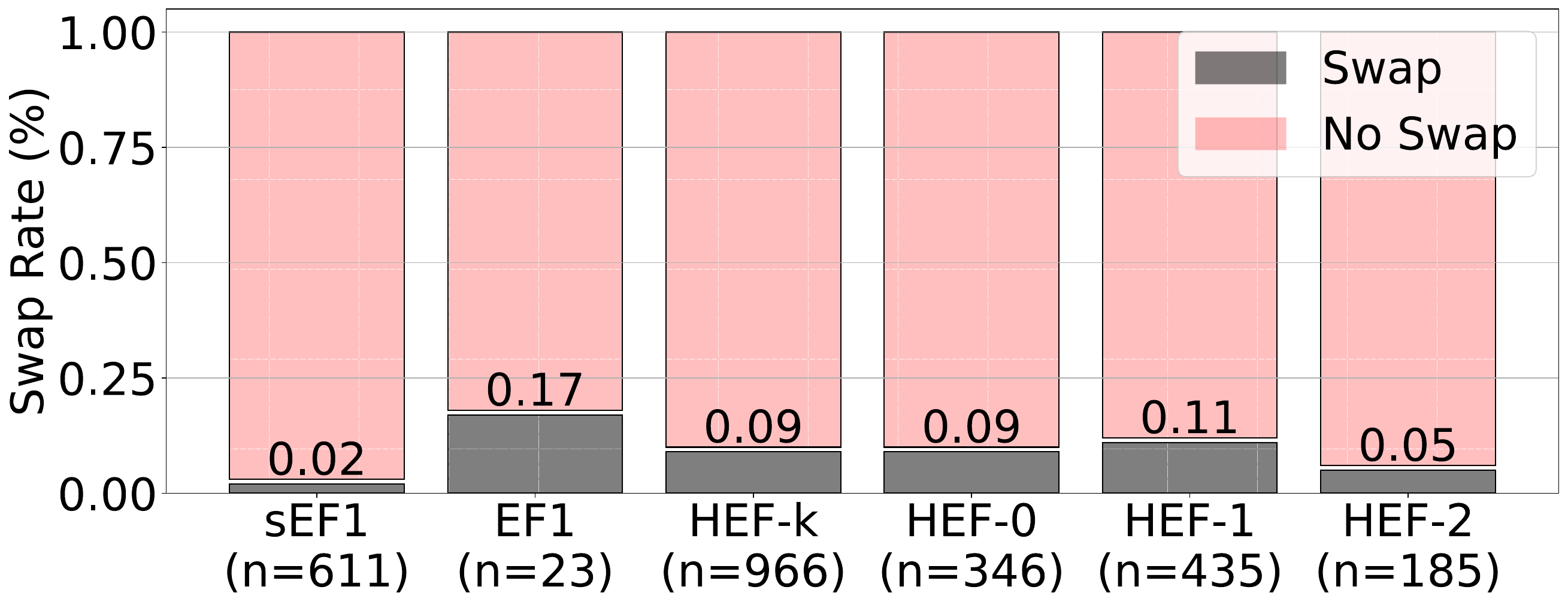}
  \caption{Swap rates per treatment, \stayopt{}. Here, $n$ is the number of questions per treatment.}
  \label{fig:mosaic-stay_is_opt}
\end{figure}

\begin{paragraph}{Optimal Choice.} We find that participants' perceived fairness is indeed affected by their optimal choice. Specifically, for each treatment (except \HEF{0}), participants' swap rates are statistically different between \swapopt{} and \stayopt{} scenarios (\cref{tab:stay_vs_swap_is_opt}).

%%%%%%%%%%%%%%%%%%%%%%%%%%%%%%%%%%%%%%%%

%%%%%%%%%%%%%%%%%%%%%%%%%%%%%%%%%%%%%%%%

Among \stayopt{} scenarios (\cref{fig:mosaic-stay_is_opt}),
% where participants' bundles have the highest value, 
we observe that \sEF{1} allocations are perceived with significantly lower envy than \HEF{k} allocations, and in turn \EF{1} allocations.
Participants of the \sEF{1} treatment could verify with certainty that their bundles have the highest value since all goods were visible. It may not be possible to make such determinations under the \HEF{1} and \HEF{2} treatments, where goods may be hidden.
% to eliminate envy between other pairs of pirates. 
Indeed, the hidden goods may all be allocated to another pirate, hypothetically raising the value of that pirate's bundle to be the highest, justifying a swap. 
Surprisingly, \HEF{0} and \EF{1} induce higher envy than \sEF{1} allocations, despite it being equally possible to verify that the participant's bundle has the highest value. 
This may be due to \emph{framing effect} biases
by which participants may not have incorrectly assumed that goods were missing \citep{tversky1985framing}.
However, further tests are needed to confirm this conjecture.
Swap rates between either \HEF{k} and \EF{1}, and \HEF{2} and \sEF{1}, are not statistically significant in this case. 
%.

When \swapopt{} (\cref{fig:mosaic-swap_is_opt} in the appendix), participants swap their bundles significantly less under the \HEF{k} treatment than the \sEF{1} and \EF{1} treatments. This supports our overall conclusion that participants desire allocations that are not visibly unfair. Since all goods are visible under the \sEF{1} and \EF{1} treatments, the participant has clear evidence that her allocated bundle has a lower value than that of another pirate. Under the \HEF{k} treatment, rather, where the allocation of some goods is hidden, participants perceive significantly lower envy even when they are allocated a lower-valued bundle. Recall that \HEF{0} is equivalent to \EF{}, so there are no such scenarios when \swapopt{}.
\end{paragraph}

\begin{paragraph}{Controlling for the choice of goods.} Our scenarios presented goods related to a pirate's adventure, such as a map, rum, and a diamond. This gamified scenario stands in for a wider variety of fair division problems, such as inheritance division \citep{Brams96:Fair}, allocating medical resources \citep{pathak2021fair}, and course allocation \citep{budish2017course}. To control for any preferential bias toward these goods,
we repeated a scenario and replaced the goods with identically-shaped gems of different colors. The repeated scenario (S7) was identical to the original (S4), which is \sunbal{} but with varying numbers of hidden goods $k$ for the \HEF{k} treatment. 
Additionally, the pictures representing the goods were randomly permuted for every scenario.

We find that every null hypothesis that was rejected by comparing responses on all scenarios is also rejected when the test is performed only on the repeated scenario (see row labeled ``Repeated scenario (S7)'' in \cref{tab:perceived_fairness}). Furthermore, the ratio of swap rates for each pair of treatments remains similar as well. Therefore, our results do not appear to be impacted by the choice of goods.
\end{paragraph}

\subsection{Cognitive Effort} \label{sec:cognitive}

In addition to our tests of perceived fairness, we investigate the extent to which cognitive effort varies by treatment. Specifically, we measure:
\begin{itemize}[leftmargin=*,topsep=0pt,itemsep=0pt]
    \item {\em response time}, the time elapsed between each scenario page being made available to the participant and the participant submitting her choice, and
    \item {\em scenario difficulty}, using the self-reports of scenario difficulty solicited immediately after the \smallq{} and then the \largeq{} scenarios.
\end{itemize}
We check whether the mean response time or reported difficulty on a five-point Likert scale is different between pairs of treatments, while adjusting for different variables such as optimal choice and instance size.
% \begin{enumerate*}[label=(\roman*)]
%     \item {\em optimal choice} -- whether the value-maximizing choice is to stay with the allocated bundle (\stayopt{}) -- and
%     \item {\em instance size} -- \smallq{} or \largeq{}.
% \end{enumerate*}

{\em Null Hypothesis:} Cognitive effort (by response time or reported difficulty) is independent of treatment.

{\em Alternate Hypothesis:} Cognitive effort differs between treatments.

\begin{figure*}[t]
  \begin{minipage}[b]{0.49\linewidth}
    \centering    \includegraphics[width=\linewidth]{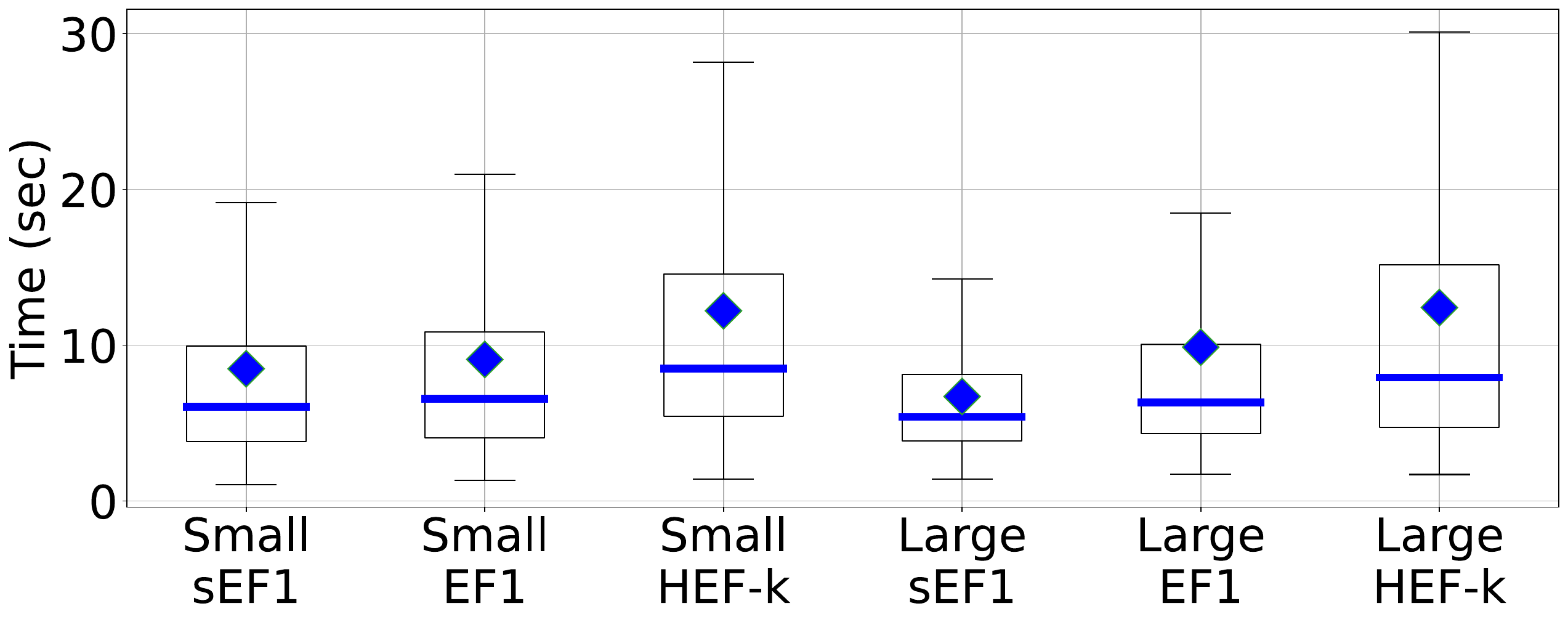}
  \end{minipage}%%
  % \hspace{1ex}
  \hfill
  \begin{minipage}[b]{0.49\linewidth}
    \centering    \includegraphics[width=\linewidth]{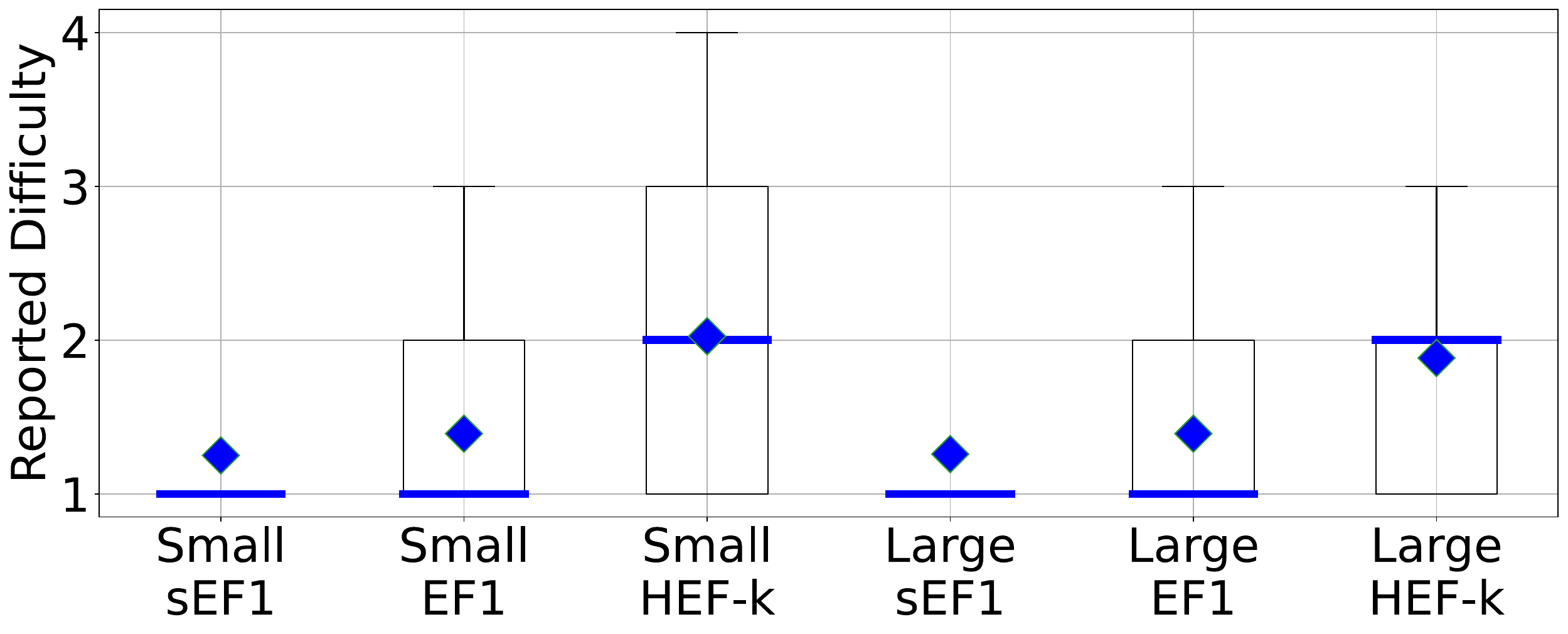}
  \end{minipage}
  \caption{Box-plots of (left) time spent per scenario and (right) reported difficulty (higher scores indicate higher difficulty) by treatment. Outliers excluded.}
  \label{fig:box_plots}
\end{figure*}

Our experiments provide sufficient evidence to reject the null hypothesis that cognitive effort for \HEF{k} is the same as either \sEF{1} or \sEF{1}, using a two-sided Welch t-test ($p<0.001$). Tables \ref{tab:times_per_treatment} and \ref{tab:feedback_per_treatment} in the appendix summarize our findings for response times per scenario and reported feedback. 
Figure~\ref{fig:box_plots} (left) illustrates that the average \sEF{1} response time is lowest and \HEF{k} is highest, while 
\EF{1} splits
the two. 
Similarly, in \cref{fig:box_plots} (right), participants report
that \sEF{1} scenarios are easiest while \HEF{k} is the most difficult and \EF{1} lies in between. 
These observations hold for either instance size and demonstrate that \HEF{k} instances cause higher cognitive burden on participants.

Note that the blue line in the middle of the box of the figures indicate the median value. The upper and lower boundaries of box show the 25th and 75th percentile respectively, and the upper and lower whiskers show the range of recorded values. The mean is indicated by a blue diamond. Outliers beyond the whiskers are excluded.
Effect size for these statistical tests, as measured by Cohen's D~\citep{Cohen92:Power}, is reported in Tables \ref{tab:times_per_treatment_Cohen_D} and \ref{tab:feedback_per_treatment_Cohen_D} in the appendix.

\subsection{Descriptive Comments from Participants}
\label{sec:comments}

We identify the participants anonymously as $S$, $E$, or $H$ corresponding to their treatment (\sEF{1}, \EF{1}, or \HEF{k}).

Participants in the \EF{1} treatment consistently noted that other pirates' bundles {\em ``were usually more valuable''} ($E22$), so they should {\em ``swap with the highest yielding chest''} ($E17$, $E8$, $E15$). On the other hand, \HEF{k} participants noted {\em ``it seemed a no brainer to just never swap''} ($H8$, $H28$), either because it was the {\em ``safest bet''} ($H49$) or the {\em ``greatest statistical chance of getting higher reward''} ($H59$).
These comments are consistent with our data that swap rates were significantly lower for \HEF{k} than the other treatments.

A few participants explicitly addressed concerns about fairness. Participant $S57$ suggested {\em ``it didn't seem like a fair split''} while $S63$ declared they wouldn't swap in real life {\em ``because it would be unfair to the other person.''} Despite this hesitation, participant $E96$ reasoned that because {\em ``there was no defining reason why anyone would get more than others''} due to differing effort, they should still select the most valuable treasure. These comments resemble \citet{herreiner2009envy}'s findings that people care more about inequality aversion than \EF{} to ensure fairness. 
Still, it is unclear to what extent participants' choices are affected by strategic interaction with other humans, as in \citet{herreiner2009envy}, as opposed to inanimate agents, as in our work. We leave this question for future work.

\section{Limitations and Future Work}

Our experiment was limited, in part, by the scenario size, uncontrolled bias, and the type of fairness notions we tested. First, our experiment tested scenarios for a cross-section of the numbers of goods $m$, agents $n$, and goods hidden $k$. We sought
to 
provide meaningful information to participants without causing cognitive overload. Future work may determine how sensitive our results are to scaling these values. 

% \HH{discuss 
% 1) other approaches to measure perceived fairness; 
% 2) impact of disparity (envy amount) on perceived fairness}

Second, we controlled for effects of our pirate-related goods on participant decision-making by randomly permuting good images and repeating a scenario with identical multi-colored gems. Still, we may not have accounted for all confounding variables, such as framing effects. For example,
while participant values were \emph{subjective}, they could have reasoned that values were \emph{objective} based on the scenario appearance. 
Furthermore, the \HEF{k} treatment conveyed to participants the possibility that hidden goods may not be allocated at all. This subsumes the reality that all goods were indeed allocated, yet is par to the definition of \HEF{k} \citep{Hosseini2020:Fair}.
This is not be the only way to implement an information scheme, as exemplified by \citet{herreiner2009envy}, who provided all subjective value information for all agents.
Further work may be necessary to determine the sensitivity of our results to framing effects and what information is provided.

Finally, our experiments compared the relative perceived fairness of two intrapersonal envy-based concepts. Both \EF{1} and \HEF{k}
presume people find their bundle fair if they 
are not envious of others' bundles; we measured 
perceived envy via swap rate according to this standard. Our results confirm that people experience less envy among allocations for which they theoretically and epistemically should not experience envy (\HEF{k}) than those requiring counterfactual reasoning (\EF{1}).
Future work could evaluate the sensitivity of our results to other measures of perceived fairness, such as the degree of envy participants perceive rather than only the binary indication of their envy.
Furthermore, whether envy-based notions are more appropriate than comparative forms, such as inequality aversion, is a topic of ongoing debate.
% \citep{herreiner2009envy}. 

Our work presents an important first step to provide an empirical comparison about perceived fairness
using the canonical envy-based definition of fairness.
% of relaxations of \EF{}.
Future empirical research may investigate perceived fairness of other notions, such as maximin-share \citep{B11combinatorial} and proportionality, attitudes towards procedural versus distributive fairness, and whether moral judgments are affected by the stake participants have in the decision problem: whether they receive resources depending on their choice or make decisions as outside observers.
% either having a stake in the resource division task or making decisions as outside observers.

\section*{Acknowledgements}

We thank the anonymous reviewers for helpful comments. HH acknowledges support from NSF grants
\#2144413, \#2107173, and \#2052488. RV acknowledges support from DST INSPIRE grant no.
\newline DST/INSPIRE/04/2020/000107. LX acknowledges support from NSF grants \#1453542, \#2007994, and
\#2106983, and a Google Research Award.

%%% Use this command to include your bibliography file.

\clearpage

\bibliographystyle{plainnat}
\bibliography{arxiv_main}

\clearpage

\appendix

\section*{Appendix}

% \section{Experimental Design: Additional Details}
% \label{apx:additional_details}

% The workflow for a participant in our study is illustrated in \cref{fig:turker_workflow}. Here, we provide details on the measures used in our study to ensure high quality responses from participants recruited from the Amazon Mechanical Turk platform. In addition to screening participants and qualifying responses, we discuss how participants were incentivized to provide high quality responses through our payments structure.

\begin{figure}[h]
  \centering %[htp]
    \includegraphics[width=\linewidth]{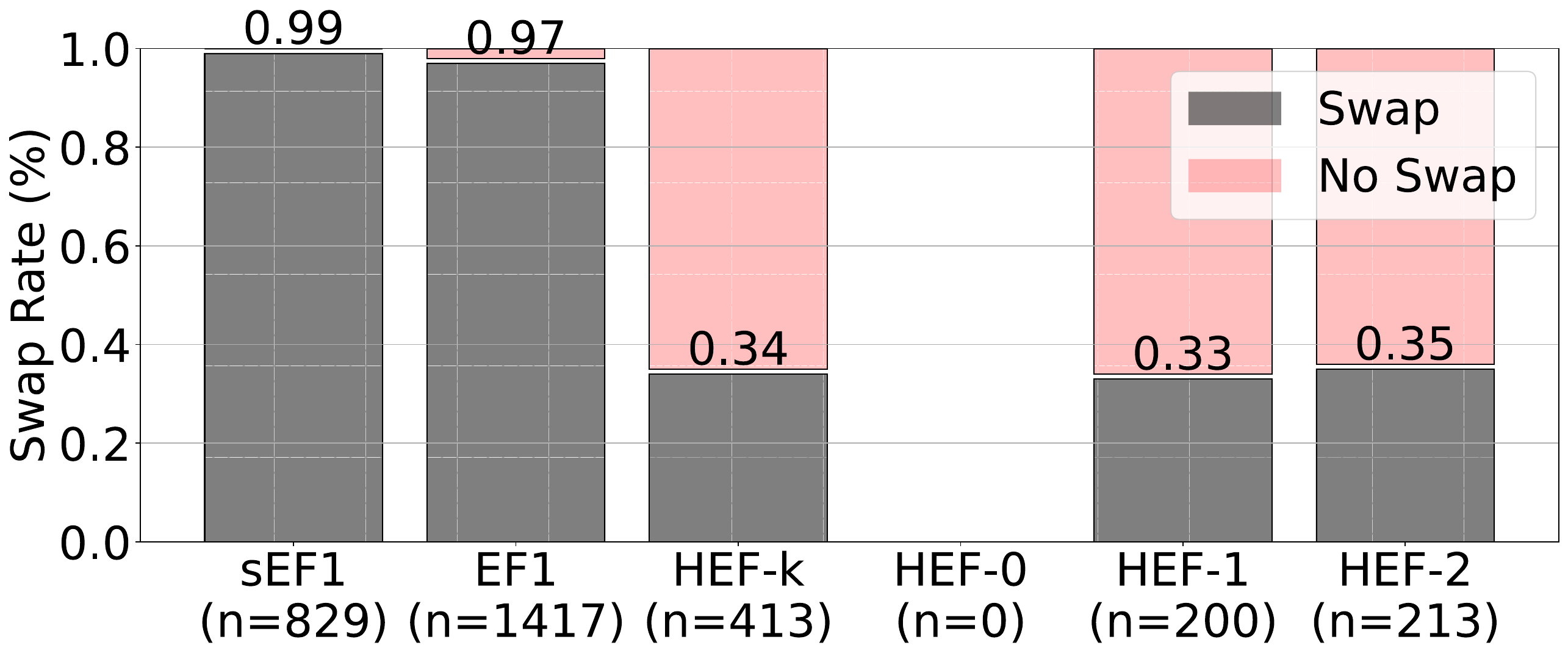}
  \caption{Swap rates per treatment, \swapopt{}. Here, $n$ is the number of questions per treatment.}
  \label{fig:mosaic-swap_is_opt}
    % \Description{Swap rates per treatment, swap-is-opt.}
\end{figure}

\begin{table}[h]
    % \tiny
    \footnotesize{}
    \centering
    \caption{
    Number of scenarios per treatment and perspective, given instance size and allocation balance (number of goods per each agent in parentheses).
    }
    \label{tab:survey_question_properties}
    \begin{tabular}{|c|c||c|c|c|c|c|}\hline
        \multirow{2}{*}{\bf \makecell{Instance\\Size}} & \multirow{2}{*}{\bf \makecell{Allocation\\Balance}} & \multicolumn{5}{c|}{\bf Treatment} \\ [0.5ex] \cline{3-7}
         & & \sEF{1} & \EF{1} & \HEF{0} & \HEF{1} & \HEF{2} \\  [0.5ex] 
         \hline \hline
        %%%%%%%%%%%%%%%%%%%%%%%%%%%%%%%%%%%%%%%%%%%%
        \multirow{2}{*}{\smallq{}}
        & \makecell{\bal{}\\$(3,3,3)$} & 
        $15$ & 
        $15$ & 
        $5$ & 
        $5$ & 
        $5$ \\ [0.5ex] 
        \cline{2-7}
        & \makecell{\unbal{}\\$(2,2,4)$} & 
        $6$ & 
        $6$ & 
        $2$ & 
        $2$ & 
        $2$ \\ [0.5ex] 
        \hline
        % %%%%%%%%%%%%%%%%%%%%%%%%%%%%%%%%%%%%%%%%%%%
        \multirow{3}{*}{\largeq{}} 
        & \makecell{\bal{}\\$(2,2,2,2,2)$} &
        $5$ & 
        $5$ & 
        $1$ & 
        $2$ & 
        $1$ \\ [0.5ex] 
        \cline{2-7}
         & \makecell{\unbal{}\\$(4,2,2,1,1)$} &
        $1$ & 
        $1$ & 
        $0$ & 
        $0$ & 
        $0$ \\ [0.5ex] 
        \cline{2-7}
        & \makecell{\unbal{}\\$(3,2,2,2,1)$} &
        $1$ & 
        $1$ & 
        $0$ & 
        $2$ & 
        $0$ \\ [0.5ex] 
        \hline
        %%%%%%%%%%%%%%%%%%%%%%%%%%%%%%%%%%%%%%%%
    \end{tabular}
\end{table}

\begin{table}[h]
    % \tiny
    \footnotesize{}
    \centering
    \caption{
    Number of participants satisfying each qualification range, per treatment, as measured by minimum approval rate and minimum approval range (not mutually exclusive).
    }
    \label{tab:response_qualifications}
    \begin{tabular}{|c|c|c|c|}\hline
        Treatment & \makecell{Minimum\\Approval Rate} & \makecell{Minimum\\Number Approved} & Count \\  [0.5ex] 
         \hline \hline
        %%%%%%%%%%%%%%%%%%%%%%%%%%%%%%%%%%%%%%%%%%%%
        \sEF{1} & $95\%$ & $1000$ & $120$ \\ [0.5ex]
        \hline \hline
        % %%%%%%%%%%%%%%%%%%%%%%%%%%%%%%%%%%%%%%%%%%%
        \multirow{2}{*}{\EF{1}} 
        & $95\%$ &
        $1000$ & 
        $20$ \\ [0.5ex] %20
        \cline{2-4}
        & $80\%$ &
        $100$ & 
        $120$ \\ [0.5ex] %100
        \hline \hline
        %%%%%%%%%%%%%%%%%%%%%%%%%%%%%%%%%%%%%%%%
        \multirow{6}{*}{\HEF{k}} 
        & $95\%$ &
        $1000$ & 
        $20$ \\ [0.5ex] %20
        \cline{2-4}
        & $90\%$ &
        $1000$ & 
        $62$ \\ [0.5ex] %42
        \cline{2-4}
        & $80\%$ &
        $1000$ & 
        $76$ \\ [0.5ex] %14
        \cline{2-4}
        & $90\%$ &
        $500$ & 
        $83$ \\ [0.5ex] %21
        \cline{2-4}
        & $90\%$ &
        $100$ & 
        $92$ \\ [0.5ex] %9
        \cline{2-4}
        & $80\%$ &
        $100$ & 
        $120$ \\ [0.5ex] %14
        \hline
        %%%%%%%%%%%%%%%%%%%%%%%%%%%%%%%%%%%%%%%%
    \end{tabular}
\end{table}

% \section{Experimental Results: Additional Tables and Figures}
% \label{apx:additional_figures}

% \paragraph{Perceived envy under different treatments.}

% \cref{tab:perceived_fairness} presents the ratio of swap rates and p-values for all questions under different pairs of treatments.

% \paragraph{Perceived envy of \swapopt{} versus \stayopt{} scenarios under different treatments.} \cref{tab:stay_vs_swap_is_opt} depicts the ratio of swap rates for \swapopt{} versus \stayopt{} scenarios under different treatments. Notably, there is a statistically significant difference between \swapopt{} and \stayopt{} for all treatments (except for \HEF{0}, for which there are no \swapopt{} scenarios). 

\paragraph{Perceived envy comparing treatments.}
First, Tables \ref{tab:perceived_fairness_full} and \ref{tab:stay_vs_swap_is_opt_full} depicts the results of hypothesis tests comparing the independence of swap rates and treatments, while adjusting for different variables. These provide more information than Tables \ref{tab:perceived_fairness} and \ref{tab:stay_vs_swap_is_opt} in the main text.

Second, \cref{tab:hef_k_perceived_fairness} presents tests for independence among the pairwise treatments of \HEF{0}, \HEF{1}, and \HEF{2}.
% , and adjusting for different variables. 
% These results complement \cref{tab:perceived_fairness} and demonstrate $\chi^2$ statistics and associated $p$-values. 
Notably, there is a statistically significant difference between \HEF{0} and both \HEF{1} and \HEF{2} for all questions, although there is no significant difference between the treatments conditioning on either \stayopt{} or \swapopt{}. By \cref{fig:mosaic-all_qs}, this suggests \HEF{0} (i.e., envy-free) allocations are perceived as more fair than either \HEF{1} or \HEF{2} allocations.

% \textcolor{red}{TODO Talk about Tables \ref{tab:perceived_fairness_full} and  \ref{tab:stay_vs_swap_is_opt_full}}

\begin{table*}[t]
    \footnotesize{}
    % \tiny
    \centering
    \caption{ Ratio of the swap rates and $p$-values of the test statistic for testing the independence of swap rates and treatments under different pairs of treatments, and adjusting for different variables. The $\Chi^2$ test is used except when the $p$-value is annotated with a ``$\dagger$'', in which case, it is the result of the Fisher's exact test. The $p$-value of the test statistic is represented as follows: a cell labeled \ns{} (not significant) implies that $p >0.05$, $\medstar$ for $p \in (0.01, 0.05]$, $\medstar \medstar$ for $p \in (0.001, 0.01]$, and $\medstar \medstar \medstar$ for $p < 0.001$. 
    % \SKS{Discuss behavior in \stayopt{}.}
    }
    % \SKS{Computed from data in `data/ContingencyTables.xlsx'} \SKS{Missing some statistical tests.}}
    % \ding{80} : $V \leq 0.2$), (\ding{80} \ding{80} : $V \in (0.2, 0.6]$), (\ding{80} \ding{80} \ding{80} : $V > 0.6$)
    \label{tab:perceived_fairness_full}
    % \label{tab:perceived_fairness}
    \begin{tabular}{|c|c||c|c|c||c|c|c|}\hline
        \multirow{2}{*}{\bf Variable} & \multirow{2}{*}{\bf value} & \multicolumn{6}{c|}{\bf Pairs of Treatments} \\ \cline{3-8}
         & & \HEF{k}, \sEF{1} & \HEF{k}, \EF{1} & \sEF{1}, \EF{1} & \HEF{0}, \sEF{1} & \HEF{1}, \sEF{1} & \HEF{2}, \sEF{1} \\ \hline \hline
        %%%%%%%%%%%%%%%%%%%%%%%%%%%%%%%%%%%%%%
        \multicolumn{2}{|c||}{All scenarios} & 
        \makecell{$0.286$\\\footnotesize{$p: \medstar \medstar \medstar$}} & 
        \makecell{$0.173$\\\footnotesize{$p: \medstar \medstar \medstar$}} & 
        \makecell{$0.604$\\\footnotesize{$p: \medstar \medstar \medstar$}} & 
        \makecell{$0.150$\\\footnotesize{$p: \medstar \medstar \medstar$}} & 
        \makecell{$0.306$\\\footnotesize{$p: \medstar \medstar \medstar$}} & 
        \makecell{$0.371$\\\footnotesize{$p: \medstar \medstar \medstar$}} \\
        \hline
        %%%%%%%%%%%%%%%%%%%%%%%%%%%%%%%%%%%%%%%%%%%%
        \multirow{2}{*}{\makecell{Optimal\\ Choice}}
        & \stayopt{} & 
        \makecell{$4.533$\\\footnotesize{$p: \medstar \medstar \medstar$}} & 
        \makecell{$0.512$\\\footnotesize{$p:\ns{}$}} & % $p=0.15$
        \makecell{$0.113$\\\footnotesize{$\dagger p: \medstar \medstar$}} & 
        \makecell{$4.415$\\\footnotesize{$p: \medstar \medstar \medstar$}} & 
        \makecell{$5.384$\\\footnotesize{$p: \medstar \medstar \medstar$}} & 
        \makecell{$2.752$\\\footnotesize{$p:\ns{}$}} \\
        \cline{2-8}
        & \swapopt & 
        \makecell{$0.346$\\\footnotesize{$p: \medstar \medstar \medstar$}} & 
        \makecell{$0.353$\\\footnotesize{$p: \medstar \medstar \medstar$}} & 
        \makecell{$1.021$\\\footnotesize{$p: \medstar$}} & 
        \makecell{\texttt{N/A}} & 
        \makecell{$0.334$\\\footnotesize{$p: \medstar \medstar \medstar$}} & 
        \makecell{$0.357$\\\footnotesize{$p: \medstar \medstar \medstar$}} \\
        \hline
        %%%%%%%%%%%%%%%%%%%%%%%%%%%%%%%%%%%%%%%%%%%
        \multirow{2}{*}{\makecell{Instance\\ Size}} 
        & \smallq{} &
        \makecell{$0.294$\\\footnotesize{$p: \medstar \medstar \medstar$}} & 
        \makecell{$0.186$\\\footnotesize{$p: \medstar \medstar \medstar$}} & 
        \makecell{$0.634$\\\footnotesize{$p: \medstar \medstar \medstar$}} & 
        \makecell{$0.114$\\\footnotesize{$p: \medstar \medstar \medstar$}} & 
        \makecell{$0.380$\\\footnotesize{$p: \medstar \medstar \medstar$}} & 
        \makecell{$0.394$\\\footnotesize{$p: \medstar \medstar \medstar$}} \\
        \cline{2-8}
        & \largeq{} &
        \makecell{$0.268$\\\footnotesize{$p: \medstar \medstar \medstar$}} & 
        \makecell{$0.150$\\\footnotesize{$p: \medstar \medstar \medstar$}} & 
        \makecell{$0.561$\\\footnotesize{$p: \medstar \medstar \medstar$}} & 
        \makecell{$0.322$\\\footnotesize{$p: \medstar \medstar \medstar$}} & 
        \makecell{$0.253$\\\footnotesize{$p: \medstar \medstar \medstar$}} & 
        \makecell{$0.285$\\\footnotesize{$p: \medstar \medstar \medstar$}} \\
        \hline
        %%%%%%%%%%%%%%%%%%%%%%%%
        \multirow{2}{*}{Balance}
        & \bal{} &
        \makecell{$0.320$\\\footnotesize{$p: \medstar \medstar \medstar$}} & 
        \makecell{$0.167$\\\footnotesize{$p: \medstar \medstar \medstar$}} & 
        \makecell{$0.523$\\\footnotesize{$p: \medstar \medstar \medstar$}} & 
        \makecell{$0.246$\\\footnotesize{$p: \medstar \medstar \medstar$}} & 
        \makecell{$0.267$\\\footnotesize{$p: \medstar \medstar \medstar$}} & 
        \makecell{$0.426$\\\footnotesize{$p: \medstar \medstar \medstar$}} \\
        \cline{2-8}
        & \unbal{} & 
        \makecell{$0.259$\\\footnotesize{$p: \medstar \medstar \medstar$}} & 
        \makecell{$0.178$\\\footnotesize{$p: \medstar \medstar \medstar$}} & 
        \makecell{$0.686$\\\footnotesize{$p: \medstar \medstar \medstar$}} & 
        \makecell{$0.073$\\\footnotesize{$p: \medstar \medstar \medstar$}} & 
        \makecell{$0.310$\\\footnotesize{$p: \medstar \medstar \medstar$}} & 
        \makecell{$0.329$\\\footnotesize{$p: \medstar \medstar \medstar$}} \\
        \hline
        %%%%%%%%%%%%%%%%%%%%%%%%%%%%%%%%%%%%%%%%
        \multicolumn{2}{|c||}{Repeated scenario (S7)} & 
        \makecell{$0.286$\\\footnotesize{$p: \medstar \medstar \medstar$}} & 
        \makecell{$0.211$\\\footnotesize{$p: \medstar \medstar \medstar$}} & 
        \makecell{$0.737$\\\footnotesize{$p: \medstar \medstar \medstar$}} & 
        \makecell{$0.091$\\\footnotesize{$p: \medstar \medstar \medstar$}} & 
        \makecell{$0.490$\\\footnotesize{$p: \medstar \medstar$}} & 
        \makecell{$0.338$\\\footnotesize{$p: \medstar \medstar \medstar$}} \\
        \hline
    \end{tabular}
\end{table*}

\begin{table*}[t] %[htp]
    % \tiny
    % \footnotesize{}
    \small
    \centering
    \caption{ Ratio of the swap rates and $p$-values of the test statistic for testing the independence of swap rates and optimal choice under different treatments. The $\Chi^2$ test is used except when the $p$-value is annotated with a ``$\dagger$'', in which case, it is the result of the Fisher's exact test. The $p$-value of the test statistic is represented as follows: a cell labeled \ns{} (not significant) implies that $p >0.05$, $\medstar$ for $p \in (0.01, 0.05]$), $\medstar \medstar$ for $p \in (0.001, 0.01]$, and $\medstar \medstar \medstar$ for $p < 0.001$.}
    \label{tab:stay_vs_swap_is_opt_full}
    \begin{tabular}{|c||c|c|c|c|c|c|}\hline
        \textbf{Treatment} & \sEF{1} & \HEF{k} & \EF{1} & \HEF{0} & \HEF{1} & \HEF{2} \\
         \hline
        \makecell{\swapopt{}\\/ \stayopt{}} & \makecell{$50.241$\\{$p: \medstar \medstar \medstar$}} & \makecell{$3.835$\\{$p: \medstar \medstar \medstar$}} & \makecell{$5.559$\\{$\dagger p: \medstar \medstar \medstar$}} & \texttt{N/A} & \makecell{$3.121$\\{$p: \medstar \medstar \medstar$}} & \makecell{$6.514$\\{$p: \medstar \medstar \medstar$}} \\ 
         \hline
    \end{tabular}
\end{table*}

\begin{figure}[t]
  \centering %[htp]
    \includegraphics[width=\linewidth]{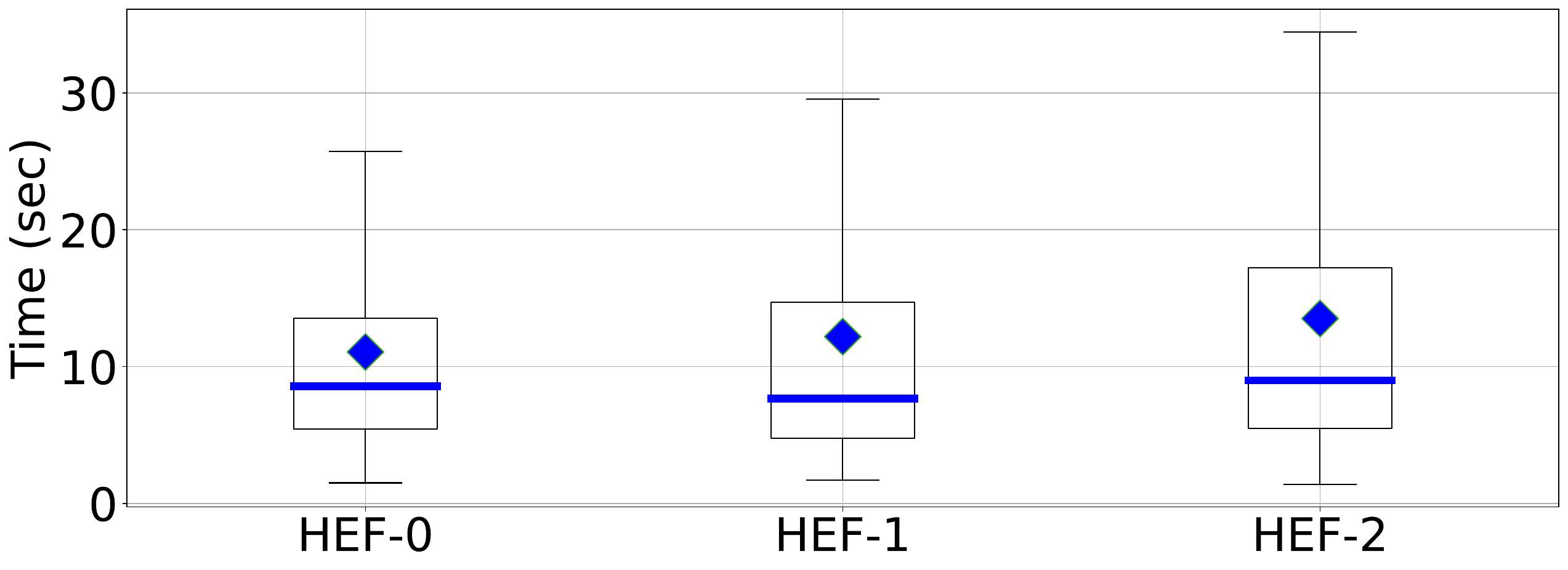}
  \caption{ Box-plot of time spent per scenario by treatment with averages shown. Outliers excluded. }
  \label{fig:plot-times-2g}
    % \Description{Box-plot of time spent per scenario by treatment with averages shown.}
\end{figure}

\begin{figure}[t]
  \centering %[htp]
  \includegraphics[width=\linewidth]{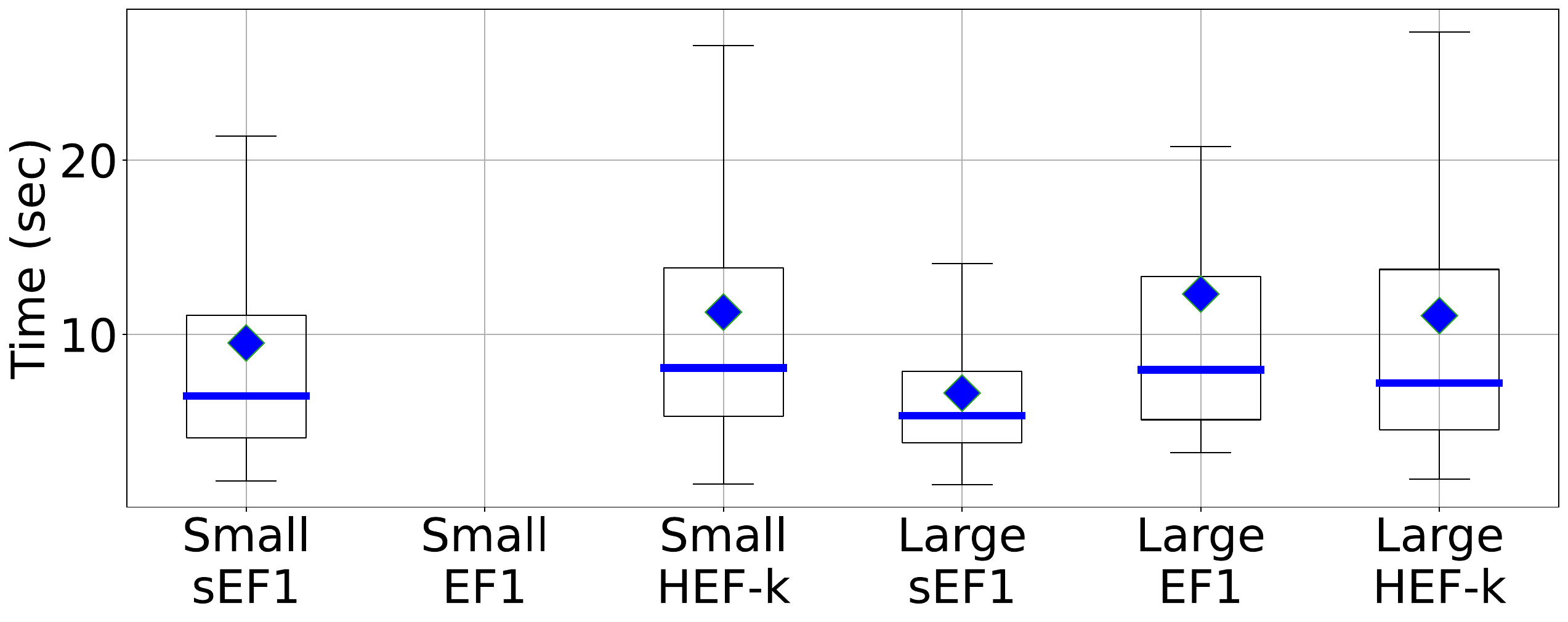}
    \caption{ Box-plot of time spent per scenario by treatment with average shown, conditioned on \stayopt{}. Outliers excluded. There were no such \smallq{} \EF{1} scenarios.}
  \label{fig:plot-times-f}
    % \Description{Box-plot of time spent per scenario by treatment with averages shown, conditioned on stay-is-opt.}
\end{figure}

\begin{table*}[t] %[htp]
    % \tiny
    \small
    \centering
    \caption{Ratio of swap rates and  $p$-values of the $\chi^2$ statistic for testing the independence of swap rates and treatments under different pairs of treatments, and adjusting for different variables. \\ Key: (\ns{} : $p >0.05$) ($\medstar$ : $p \in (0.01, 0.05]$), ($\medstar \medstar$ : $p \in (0.001, 0.01]$), ($\medstar \medstar \medstar$ : $p < 0.001$). }
    \label{tab:hef_k_perceived_fairness}
    \begin{tabular}{|c|c||c|c|c|}\hline
        \multirow{2}{*}{\bf Variable} & \multirow{2}{*}{\bf value} & \multicolumn{3}{c|}{\bf Pairs of Treatments} \\ \cline{3-5}
         & & \HEF{0}, \HEF{1} & \HEF{0}, \HEF{2} & \HEF{1}, \HEF{2} \\ \hline \hline
        %%%%%%%%%%%%%%%%%%%%%%%%%%%%%%%%%%%%%%
        \multicolumn{2}{|c||}{All scenarios} & 
        \makecell{$0.494$\\{$p: \medstar \medstar$}} & 
        \makecell{$0.406$\\{$p: \medstar \medstar \medstar$}} & 
        \makecell{$0.826$\\{$p:\ns{}$}} \\
        \hline
        %%%%%%%%%%%%%%%%%%%%%%%%%%%%%%%%%%%%%%%%%%%%
        \multirow{2}{*}{\makecell{Optimal\\ Choice}}
        & \stayopt{} & 
        \makecell{$0.820$\\{$p:\ns{}$}} & 
        \makecell{$1.604$\\{$p:\ns{}$}} & 
        \makecell{$1.956$\\{$p:\ns{}$}} \\
        \cline{2-5}
        & \swapopt & 
        % \makecell{$\text{N/A}$\\{$p=?$}} & 
        % \makecell{$\text{N/A}$\\{$p=?$}} & 
        \makecell{$\text{N/A}$} & 
        \makecell{$\text{N/A}$} & 
        \makecell{$0.937$\\{$p:\ns{}$}} \\
        \hline
        %%%%%%%%%%%%%%%%%%%%%%%%%%%%%%%%%%%%%%%%%%%
        \multirow{2}{*}{\makecell{Instance\\ Size}} 
        & \smallq{} &
        \makecell{$0.299$\\{$p: \medstar \medstar \medstar$}} & 
        \makecell{$0.289$\\{$p: \medstar \medstar \medstar$}} & 
        \makecell{$0.966$\\{$p: \ns{}$}} \\
        \cline{2-5}
        & \largeq{} &
        \makecell{$1.271$\\{$p:\ns{}$}} & 
        \makecell{$1.130$\\{$p:\ns{}$}} & 
        \makecell{$0.889$\\{$p:\ns{}$}} \\
        \hline
        %%%%%%%%%%%%%%%%%%%%%%%%
        \multirow{2}{*}{Balance}
        & \bal{} &
        \makecell{$0.992$\\{$p:\ns{}$}} & 
        \makecell{$0.578$\\{$p:\ns{}$}} & 
        \makecell{$0.627$\\{$p:\ns{}$}} \\
        \cline{2-5}
        & \unbal{} & 
        \makecell{$0.237$\\{$p: \medstar \medstar \medstar$}} & 
        \makecell{$0.223$\\{$p: \medstar \medstar \medstar$}} & 
        \makecell{$0.941$\\{$p:\ns{}$}} \\ 
        \hline
        %%%%%%%%%%%%%%%%%%%%%%%%%%%%%%%%%%%%%%%%
        \multicolumn{2}{|c||}{Repeated scenario (Q7)} & 
        \makecell{$0.186$\\{$p: \medstar$}} & 
        \makecell{$0.270$\\{$p: \ns{}$}} & 
        \makecell{$1.448$\\{$p:\ns{}$}} \\
        \hline
    \end{tabular}
\end{table*}

\begin{table*}[t] %[htp]
    % \tiny
    \small
    \centering
    \caption{ $p$-values of the $t$ statistic for testing equal means of participant response times per scenario using Welch's t-test -- for different pairs of treatments, and adjusting for different variables. \\ Key: (\ns{} : $p >0.05$) ($\medstar$ : $p \in (0.01, 0.05]$), ($\medstar \medstar$ : $p \in (0.001, 0.01]$), ($\medstar \medstar \medstar$ : $p < 0.001$). }
    \label{tab:times_per_treatment}
    \begin{tabular}{|c|c||c|c|c|}\hline
        \multirow{2}{*}{\bf Variable} & \multirow{2}{*}{\bf Instance Size} & \multicolumn{3}{c|}{\bf Pairs of Treatments} \\ \cline{3-5}
         & & \sEF{1}, \EF{1} & \sEF{1}, \HEF{k} & \EF{1}, \HEF{k} \\ \hline \hline
        %%%%%%%%%%%
        %%%%%%%%%%%%%%%%%%%%%%%%%%%
        \multirow{2}{*}{\makecell{All scenarios}}
        & \smallq{} & 
        \makecell{{$p:\ns{}$}} & 
        \makecell{{$p:\medstar \medstar \medstar$}} & 
        \makecell{{$p:\medstar \medstar \medstar$}} \\
        \cline{2-5}
        & \largeq & 
        \makecell{{$p:\medstar$}} & % changed from 3* to 1* (4/16/23)
        \makecell{{$p:\medstar \medstar \medstar$}} & 
        \makecell{{$p:\medstar \medstar \medstar$}} \\
        \hline
        %%%%%%%%%%%%%%%%%%%%%%%%%%%%%%%%%%%%%%%%%%%
        \multirow{2}{*}{\makecell{\stayopt{}}} 
        & \smallq{} &
        \makecell{{$\text{N/A}$}} & 
        \makecell{{$p:\medstar$}} & 
        \makecell{{$\text{N/A}$}} \\
        \cline{2-5}
        & \largeq{} &
        \makecell{{$p:\ns{}$}} & 
        \makecell{{$p:\medstar \medstar \medstar$}} & 
        \makecell{{$p:\ns{}$}} \\
        \hline \hline
        %% SPLIT %%
        \multirow{2}{*}{\bf Variable} & \multirow{2}{*}{} & \multicolumn{3}{c|}{\bf Pairs of Treatments} \\ \cline{3-5}
         & & \HEF{0}, \HEF{1} & \HEF{0}, \HEF{2} & \HEF{1}, \HEF{2} \\ \hline \hline
        %%%%%%%%%%%%%%%%%%%%%%%%%%%%%%%%%%%%%%
        All scenarios
        &  & 
        \makecell{{$p:\ns{}$}} & 
        \makecell{{$p:\medstar \medstar$}} & 
        \makecell{{$p:\ns{}$}} \\
        \hline
    \end{tabular}
\end{table*}

\begin{table*}[t] %[htp]
    % \tiny
    \small
    \centering
    \caption{$p$-values of the $t$ statistic for testing equal means of participant reported feedback using Welch's t-test --  for different pairs of treatments. \\ Key: (\ns{} : $p >0.05$) ($\medstar$ : $p \in (0.01, 0.05]$), ($\medstar \medstar$ : $p \in (0.001, 0.01]$), ($\medstar \medstar \medstar$ : $p < 0.001$). }
    \label{tab:feedback_per_treatment}
    \begin{tabular}{|c|c||c|c|c|}\hline
        \multirow{2}{*}{\bf Variable} & \multirow{2}{*}{\bf Instance Size} & \multicolumn{3}{c|}{\bf Pairs of Treatments} \\ \cline{3-5}
         & & \sEF{1}, \EF{1} & \sEF{1}, \HEF{k} & \EF{1}, \HEF{k} \\ \hline \hline
        %%%%%%%%%%%%%%%%%%%%%%%%%%%%%%%%%%%%%%
        \multirow{2}{*}{\makecell{All senarios}}
        & \smallq{} & 
        \makecell{{$p:\medstar$}} & 
        \makecell{{$p:\medstar \medstar \medstar$}} & 
        \makecell{{$p:\medstar \medstar \medstar$}} \\
        \cline{2-5}
        & \largeq & 
        \makecell{{$p:\ns{}$}} & 
        \makecell{{$p:\medstar \medstar \medstar$}} & 
        \makecell{{$p:\medstar \medstar \medstar$}} \\
        \hline
    \end{tabular}
\end{table*}

\paragraph{Cognitive effort on \HEF{k} allocations.}

\cref{fig:plot-times-2g} presents the distribution of time spent per scenario over all \HEF{0}, \HEF{1} and \HEF{2} scenarios. We find that overall, as the number of hidden goods increases, the cognitive effort, measured as the amount of time spent in order to decide which bundle to keep, also increases. Specifically, both the mean and variance of time spent increases as the value of $k$ increases for \HEF{k} scenarios. 

Notice that in an \HEF{0} scenario, the participant already has the highest valued bundle and this is readily verifiable since all goods are visible. However, as $k$ increases, the participant must reason about and form beliefs about how the hidden goods may be allocated to the other pirates. The task of computing and deciding whether it may be worth swapping for another pirate's bundle therefore becomes increasingly more complex as more goods are hidden.

\paragraph{Cognitive effort conditioned on \stayopt{} scenarios.} As \cref{fig:plot-times-f} shows for \stayopt{} scenarios, hiding goods under the \HEF{k} treatment comes at the cost of an increased cognitive burden on the participants. Here, the participant's bundle has the highest value. This is evident for the \sEF{1} and \EF{1} treatments, but may not be clear under the \HEF{k} treatment, where goods may need to be hidden in order to eliminate envy between the other pirates.

%%%%%%%%%%%%%%%%%%%%%%%%%%%%%%%%%%%%%%%%%%%%%%%%%%%%%%%%%%%%%%
% DO NOT ADD FIGURES ABOVE THIS LINE
%%%%%%%%%%%%%%%%%%%%%%%%%%%%%%%%%%%%%%%%%%%%%%%%%%%%%%%%%%%%%%

\paragraph{Effect Size}

We supplement our results of statistical significance with their effect sizes. \cref{tab:perceived_fairness_cramer_v}, \cref{tab:stay_vs_swap_is_opt_Cramer_V}, \cref{tab:hef_k_perceived_fairness_Cramer_V}, \cref{tab:times_per_treatment_Cohen_D}, and \cref{tab:feedback_per_treatment_Cohen_D} demonstrate the effect size for each statistically significant test for \cref{tab:perceived_fairness_full},
\cref{tab:stay_vs_swap_is_opt_full},
\cref{tab:hef_k_perceived_fairness}, \cref{tab:times_per_treatment}, and \cref{tab:feedback_per_treatment} respectively. Effect sizes are measured with Cramer's V for $\chi^2$ tests \citep{Cramer1946:Mathematical} and Cohen's d for Welch $t$-tests \citep{Cohen92:Power}.

\begin{table*}[t] %[htp]
    \footnotesize{}
    % \tiny
    \centering
    \caption{ Effect size demonstrating the strength in statistically significant relationships between swap rates and treatments -- under different pairs of treatments, adjusting for different variables, and corresponding to tests in \cref{tab:perceived_fairness}. Not significant tests are labelled \ns{}. Cramer's V is reported for $\chi^2$ tests as follows: $\medstar$ for $V \leq 0.2$, $\medstar \medstar$ for $V \in (0.2, 0.6]$, and $\medstar \medstar \medstar$ for $V > 0.6$. Odds ratio and 95\% confidence intervals are reported for Fisher's exact test, annotated by ``$\dagger$.''}
    \label{tab:perceived_fairness_cramer_v}
    \begin{tabular}{|c|c||c|c|c||c|c|c|}\hline
        \multirow{2}{*}{\bf Variable} & \multirow{2}{*}{\bf value} & \multicolumn{6}{c|}{\bf Pairs of Treatments} \\ \cline{3-8}
         & & \HEF{k}, \sEF{1} & \HEF{k}, \EF{1} & \sEF{1}, \EF{1} & \HEF{0}, \sEF{1} & \HEF{1}, \sEF{1} & \HEF{2}, \sEF{1} \\ \hline \hline
        %%%%%%%%%%%%%%%%%%%%%%%%%%%%%%%%%%%%%%
        \multicolumn{2}{|c||}{All scenarios} & 
        \makecell{$V: \medstar \medstar$} & 
        \makecell{$V: \medstar \medstar \medstar$} &
        \makecell{$V: \medstar \medstar$} & 
        \makecell{$V: \medstar \medstar$} &
        \makecell{$V: \medstar \medstar$} &
        \makecell{$V: \medstar \medstar$} \\
        \hline
        %%%%%%%%%%%%%%%%%%%%%%%%%%%%%%%%%%%%%%%%%%%%
        \multirow{2}{*}{\makecell{Optimal\\ Choice}}
        & \stayopt{} & 
        \makecell{$V: \medstar$} & 
        \makecell{\ns{}} &
        \makecell{$\dagger OR: 0.096$\\$95\%~CI:$\\(0.026, 0.447)} & 
        \makecell{$V: \medstar$} &
        \makecell{$V: \medstar$} &
        \makecell{\ns{}} \\
        \cline{2-8}
        & \swapopt & 
        \makecell{$V: \medstar \medstar \medstar$} & 
        \makecell{$V: \medstar \medstar \medstar$} &
        \makecell{$V: \medstar$} & 
        \makecell{\texttt{N/A}} & 
        \makecell{$V: \medstar \medstar \medstar$} &
        \makecell{$V: \medstar \medstar \medstar$} \\
        \hline
        %%%%%%%%%%%%%%%%%%%%%%%%%%%%%%%%%%%%%%%%%%%
        \multirow{2}{*}{\makecell{Instance\\ Size}} 
        & \smallq{} &
        \makecell{$V: \medstar \medstar$} & 
        \makecell{$V: \medstar \medstar \medstar$} &
        \makecell{$V: \medstar \medstar$} & 
        \makecell{$V: \medstar \medstar$} &
        \makecell{$V: \medstar \medstar$} &
        \makecell{$V: \medstar \medstar$} \\
        \cline{2-8}
        & \largeq{} &
        \makecell{$V: \medstar \medstar$} & 
        \makecell{$V: \medstar \medstar \medstar$} & 
        \makecell{$V: \medstar \medstar$} & 
        \makecell{$V: \medstar \medstar$} &
        \makecell{$V: \medstar \medstar$} &
        \makecell{$V: \medstar \medstar$} \\
        \hline
        %%%%%%%%%%%%%%%%%%%%%%%%
        \multirow{2}{*}{Balance}
        & \bal{} &
        \makecell{$V: \medstar \medstar$} & 
        \makecell{$V: \medstar \medstar \medstar$} &
        \makecell{$V: \medstar \medstar$} & 
        \makecell{$V: \medstar \medstar$} &
        \makecell{$V: \medstar \medstar$} &
        \makecell{$V: \medstar \medstar$} \\
        \cline{2-8}
        & \unbal{} & 
        \makecell{$V: \medstar \medstar$} & 
        \makecell{$V: \medstar \medstar \medstar$} & 
        \makecell{$V: \medstar \medstar$} & 
        \makecell{$V: \medstar \medstar$} &
        \makecell{$V: \medstar \medstar$} &
        \makecell{$V: \medstar \medstar$} \\
        \hline
        %%%%%%%%%%%%%%%%%%%%%%%%%%%%%%%%%%%%%%%%
        \multicolumn{2}{|c||}{Repeated scenario (S7)} & 
        \makecell{$V: \medstar \medstar$} &  
        \makecell{$V: \medstar \medstar \medstar$} &
        \makecell{$V: \medstar \medstar$} & 
        \makecell{$V: \medstar \medstar$} &
        \makecell{$V: \medstar \medstar$} &
        \makecell{$V: \medstar \medstar$} \\
        \hline
    \end{tabular}
\end{table*}

\begin{table*}[t] %[htp]
    % \tiny
    % \footnotesize{}
    \small
    \centering
    \caption{ Effect size demonstrating the strength in statistically significant relationships between swap rates and optimal choice, for different treatments in \cref{tab:stay_vs_swap_is_opt}. Not significant tests are labelled \ns{}. Cramer's V is reported for $\chi^2$ tests as follows: $\medstar$ for $V \leq 0.2$, $\medstar \medstar$ for $V \in (0.2, 0.6]$, and $\medstar \medstar \medstar$ for $V > 0.6$. Odds ratio and 95\% confidence intervals are reported for Fisher's exact test, annotated by ``$\dagger$.'' }
    \label{tab:stay_vs_swap_is_opt_Cramer_V}
    \begin{tabular}{|c||c|c|c|c|c|c|}\hline
        \textbf{Treatment} & \sEF{1} & \HEF{k} & \EF{1} & \HEF{0} & \HEF{1} & \HEF{2} \\
         \hline
        \makecell{\swapopt{}\\/ \stayopt{}} & $V: \medstar \medstar \medstar$ & $V: \medstar \medstar$ & \makecell{$\dagger OR: 0.007$\\$95\%~CI: (0.002, 0.023)$} & \texttt{N/A} & $V: \medstar \medstar$ & $V: \medstar \medstar$ \\ 
         \hline
    \end{tabular}
\end{table*}

\begin{table*}[t] %[htp]
    % \tiny
    \small
    \centering
    \caption{ Effect size measured by Cramer's V for $\chi^2$ tests corresponding with \cref{tab:hef_k_perceived_fairness}, under different pairs of treatments and adjusting for different variables. Not significant tests are labelled as \ns{}. \\ Key: (\ns{} : $p >0.05$) ($\medstar$ : $V \leq 0.2$), ($\medstar \medstar$ : $p \in (0.2, 0.6]$), ($\medstar \medstar \medstar$ : $p > 0.6$). }
    \label{tab:hef_k_perceived_fairness_Cramer_V}
    \begin{tabular}{|c|c||c|c|c|}\hline
        \multirow{2}{*}{\bf Variable} & \multirow{2}{*}{\bf value} & \multicolumn{3}{c|}{\bf Pairs of Treatments} \\ \cline{3-5}
         & & \HEF{0}, \HEF{1} & \HEF{0}, \HEF{2} & \HEF{1}, \HEF{2} \\ \hline \hline
        %%%%%%%%%%%%%%%%%%%%%%%%%%%%%%%%%%%%%%
        \multicolumn{2}{|c||}{All scenarios} & 
        \makecell{$V: \medstar$} & 
        \makecell{$V: \medstar$} & 
        \makecell{\ns{}} \\
        \hline
        %%%%%%%%%%%%%%%%%%%%%%%%%%%%%%%%%%%%%%%%%%%%
        \multirow{2}{*}{\makecell{Optimal\\ Choice}}
        & \stayopt{} & 
        \makecell{\ns{}} & 
        \makecell{\ns{}} & 
        \makecell{\ns{}} \\
        \cline{2-5}
        & \swapopt & 
        % \makecell{$\text{N/A}$\\{$p=?$}} & 
        % \makecell{$\text{N/A}$\\{$p=?$}} & 
        \makecell{$\text{N/A}$} & 
        \makecell{$\text{N/A}$} & 
        \makecell{\ns{}} \\
        \hline
        %%%%%%%%%%%%%%%%%%%%%%%%%%%%%%%%%%%%%%%%%%%
        \multirow{2}{*}{\makecell{Instance\\ Size}} 
        & \smallq{} &
        \makecell{$V: \medstar \medstar$} & 
        \makecell{$V: \medstar \medstar$} & 
        \makecell{\ns{}} \\
        \cline{2-5}
        & \largeq{} &
        \makecell{\ns{}} & 
        \makecell{\ns{}} & 
        \makecell{\ns{}} \\
        \hline
        %%%%%%%%%%%%%%%%%%%%%%%%
        \multirow{2}{*}{Balance}
        & \bal{} &
        \makecell{\ns{}} & 
        \makecell{\ns{}} & 
        \makecell{\ns{}} \\
        \cline{2-5}
        & \unbal{} & 
        \makecell{$V: \medstar$} & 
        \makecell{$V: \medstar \medstar$} &  
        \makecell{\ns{}} \\ 
        \hline
        %%%%%%%%%%%%%%%%%%%%%%%%%%%%%%%%%%%%%%%%
        \multicolumn{2}{|c||}{Repeated scenario (S7)} & 
        \makecell{$V: \medstar \medstar$} & 
        \makecell{\ns{}} & 
        \makecell{\ns{}} \\
        \hline
    \end{tabular}
\end{table*}

\begin{table*}[t] %[htp]
    % \tiny
    \small
    \centering
     \caption{ Effect size measured by Cohen's d for Welch t-tests corresponding with \cref{tab:times_per_treatment}, under different pairs of treatments and adjusting for different variables. Not significant tests are labelled as \ns{}. \\ Key: ($\medstar$ : $d \leq 0.3$), ($\medstar \medstar$ : $d \in (0.3, 0.7]$), ($\medstar \medstar \medstar$ : $d > 0.7$). }
    \label{tab:times_per_treatment_Cohen_D}
    \begin{tabular}{|c|c||c|c|c|}\hline
        \multirow{2}{*}{\bf Variable} & \multirow{2}{*}{\bf Instance Size} & \multicolumn{3}{c|}{\bf Pairs of Treatments} \\ \cline{3-5}
         & & \sEF{1}, \EF{1} & \sEF{1}, \HEF{k} & \EF{1}, \HEF{k} \\ \hline \hline
        %%%%%%%%%%%%%%%%%%%%%%%%%%%%%%%%%%%%%%
        \multirow{2}{*}{\makecell{All scenarios}}
        & \smallq{} & 
        \makecell{{$\ns{}$}} & 
        \makecell{{$d: \medstar \medstar$}} & 
        \makecell{{$d: \medstar$}} \\
        \cline{2-5}
        & \largeq & 
        \makecell{{$d: \medstar \medstar$}} & % changed from 3* to 2* (4/16/23)
        \makecell{{$d: \medstar \medstar$}} & 
        \makecell{{$d: \medstar \medstar$}} \\
        \hline
        %%%%%%%%%%%%%%%%%%%%%%%%%%%%%%%%%%%%%%%%%%%
        \multirow{2}{*}{\makecell{\stayopt{}}} 
        & \smallq{} &
        \makecell{{$\text{N/A}$}} & 
        \makecell{{$d: \medstar$}} & 
        \makecell{{$\text{N/A}$}} \\
        \cline{2-5}
        & \largeq{} &
        \makecell{{$\ns{}$}} & 
        \makecell{{$d: \medstar \medstar$}} & 
        \makecell{{$\ns{}$}} \\
        \hline \hline
        %% SPLIT %%
        \multirow{2}{*}{\bf Variable} & \multirow{2}{*}{} & \multicolumn{3}{c|}{\bf Pairs of Treatments} \\ \cline{3-5}
         & & \HEF{0}, \HEF{1} & \HEF{0}, \HEF{2} & \HEF{1}, \HEF{2} \\ \hline \hline
        %%%%%%%%%%%%%%%%%%%%%%%%%%%%%%%%%%%%%%
        All scenarios
        &  & 
        \makecell{{$\ns{}$}} & 
        \makecell{{$d: \medstar$}} & 
        \makecell{{$\ns{}$}} \\
        \hline
    \end{tabular}
\end{table*}

\begin{table*}[t] %[htp]
    % \tiny
    \small
    \centering
    \caption{ Effect size measured by Cohen's d for Welch t-tests corresponding with \cref{tab:feedback_per_treatment}, under different pairs of treatments and adjusting for different variables. Not significant tests are labelled as \ns{}. \\ Key: ($\medstar$ : $d \leq 0.3$), ($\medstar \medstar$ : $d \in (0.3, 0.7]$), ($\medstar \medstar \medstar$ : $d > 0.7$). }
    \label{tab:feedback_per_treatment_Cohen_D}
    \begin{tabular}{|c|c||c|c|c|}\hline
        \multirow{2}{*}{\bf Variable} & \multirow{2}{*}{\bf Instance Size} & \multicolumn{3}{c|}{\bf Pairs of Treatments} \\ \cline{3-5}
         & & \sEF{1}, \EF{1} & \sEF{1}, \HEF{k} & \EF{1}, \HEF{k} \\ \hline \hline
        %%%%%%%%%%%%%%%%%%%%%%%%%%%%%%%%%%%%%%
        \multirow{2}{*}{\makecell{All scenarios}}
        & \smallq{} & 
        \makecell{{$d: \medstar$}} & 
        \makecell{{$d: \medstar \medstar \medstar$}} & 
        \makecell{{$d: \medstar \medstar \medstar$}} \\
        \cline{2-5}
        & \largeq & 
        \makecell{{$\ns{}$}} & 
        \makecell{{$d: \medstar \medstar \medstar$}} & 
        \makecell{{$d: \medstar \medstar \medstar$}} \\
        \hline
    \end{tabular}
\end{table*}

\end{document}